\documentclass[pdflatex,sn-mathphys-num]{sn-jnl}

\usepackage{graphicx}%
\usepackage{multirow}%
\usepackage{amsmath,amssymb,amsfonts}%
\usepackage{amsthm}%
\usepackage{mathrsfs}%
\usepackage[title]{appendix}%
\usepackage{xcolor}%
\usepackage{textcomp}%
\usepackage{manyfoot}%
\usepackage{booktabs}%
\usepackage[utf8]{inputenc}
\usepackage{algorithm}%
\usepackage[table,xcdraw]{xcolor}
\usepackage{algorithmicx}%
\usepackage{algpseudocode}%
\usepackage{listings}%
\usepackage{makecell}
\usepackage{geometry}
\usepackage[most]{tcolorbox}
\usepackage{placeins}

\newtcolorbox{promptbox}[1][]{
  enhanced,
  breakable,
  sharp corners,
  colback=white,
  colframe=black!30,
  boxrule=0.8pt,
  title=Prompt,
  colbacktitle=black!40,
  coltitle=white,
  fonttitle=\normalfont\bfseries, 
  fontupper=\normalfont,          
  left=8mm,right=8mm,top=6mm,bottom=6mm,
  #1
}

\theoremstyle{thmstyleone}%
\theoremstyle{thmstyletwo}%

\theoremstyle{thmstylethree}%

\begin{document}

\title[Article Title]{
AD-CARE: A Criteria-guided LLM Agent Framework for Alzheimer's Disease Dementia Assessment from Heterogeneous Routine Clinical Data}

\author[1]{\fnm{Wenlong} \sur{Hou}}\email{willen-wenlong.hou@connect.polyu.hk}\equalcont

\author[2]{\fnm{Sheng} \sur{Bi}}\email{bisheng19970410@icloud.com}
\equalcont{These authors contributed equally to this work.}

\author[1]{\fnm{Guangqian} \sur{Yang}}\email{guangqian.yang@connect.polyu.hk}

\author[3]{\fnm{Lihao} \sur{Liu}}\email{ll610@cam.ac.uk}

\author[1]{\fnm{Ye} \sur{Du}}\email{duyee.du@connect.polyu.hk}

\author[2]{\fnm{Hanxiao} \sur{Xue}}\email{xhx1042271243@163.com}

\author[1]{\fnm{Juncheng} \sur{Wang}}\email{wjc2830@gmail.com}

\author[4,5]{\fnm{Yuxiang} \sur{Feng}}\email{fengyx@zju.edu.cn}

\author[1]{\fnm{Yue} \sur{Xun}}\email{yue523.xun@connect.polyu.hk}

\author[1]{\fnm{Nanxi} \sur{Yu}}\email{nx-nancy.yu@connect.polyu.hk}

\author[6]{\fnm{Ning} \sur{Mao}}\email{maoning@pku.edu.cn}

\author[1]{\fnm{Mo} \sur{Yang}}\email{mo.yang@polyu.edu.hk}
\author[7]{\fnm{Jing} \sur{Qin}}\email{harry.qin@polyu.edu.hk}

\author[8]{\fnm{Ling} \sur{Long}}\email{longl3@mail.sysu.edu.cn}

\author[9]{\fnm{Kay Chen} \sur{Tan}}\email{kctan@polyu.edu.hk}


\author[10]{\fnm{Shujie} \sur{Liu}}\email{shujliu@microsoft.com}

\author[11]{\fnm{Qian} \sur{Niu}}\email{qian.niu@weblab.t.u-tokyo.ac.jp}

\author[2]{\fnm{Jie} \sur{Lu}}\email{imaginglu@hotmail.com}

\author[12]{\fnm{Zoe} \sur{Kourtzi}}\email{zk240@cam.ac.uk}

\author*[13]{\fnm{Yi Wah Eva} \sur{Cheung}}\email{ceva@hku.hk}

\author*[2]{\fnm{Shaozhen} \sur{Yan}}\email{yansz\_me@ccmu.edu.cn}

\author*[14]{\fnm{Xiaomeng} \sur{Ma}}\email{mxiaom@mail.sysu.edu.cn}

\author*[1,15,16]{\fnm{Shujun} \sur{Wang}}\email{shu-jun.wang@polyu.edu.hk}

\affil[1]{\orgdiv{Department of Biomedical Engineering}, \orgname{The Hong Kong Polytechnic University}, \orgaddress{\city{Hong Kong SAR}, \country{China}}}

\affil[2]{\orgdiv{Department of Radiology and Nuclear Medicine}, \orgname{Xuanwu Hospital, Capital Medical University}, \city{Beijing}, \country{China}}

\affil[3]{\orgname{Amazon}, \country{USA}}

\affil[4]{\orgdiv{College of Control Science and Engineering}, \orgname{Zhejiang University}, \city{Hangzhou}, \country{China}}

\affil[5]{\orgname{Sany AI}, \city{Hangzhou}, \country{China}}

\affil[6]{\orgdiv{Department of Radiology}, \orgname{Yantai Yuhuangding Hospital, Qingdao University}, \city{Yantai}, \country{China}}

\affil[7]{\orgdiv{School of Nursing}, \orgname{The Hong Kong Polytechnic University}, \orgaddress{\city{Hong Kong SAR}, \country{China}}}

\affil[8]{\orgdiv{Department of Geriatrics}, \orgname{The Third Affiliated Hospital of Sun Yat-sen University}, \city{Guangzhou}, \country{China}}


\affil[9]{\orgdiv{Department of Data Science and Artificial Intelligence}, \orgname{The Hong Kong Polytechnic University}, \orgaddress{\city{Hong Kong SAR}, \country{China}}}

\affil[10]{\orgdiv{Media Computing Group}, \orgname{Microsoft Research}, \city{Beijing}, \country{China}}

\affil[11]{\orgdiv{
Department of Technology Management for Innovation, Graduate School of Engineering}, \orgname{The University of Tokyo}, \city{Tokyo}, \country{Japan}}

\affil[12]{\orgdiv{Department of Psychology}, \orgname{University of Cambridge}, \orgaddress{\country{UK}}}

\affil[13]{\orgdiv{Department of Diagnostic Radiology, School of Clinical Medicine, LKS Faculty of Medicine}, \orgname{The University of Hong Kong}, \city{Hong Kong SAR}, \country{China}}

\affil[14]{\orgdiv{Department of Neurology, Mental and Neurological Diseases Research Center}, \orgname{The Third Affiliated Hospital of Sun Yat-sen University}, \city{Guangzhou}, \country{China}}

\affil[15]{\orgdiv{Research Institute for Smart Ageing}, \orgname{The Hong Kong Polytechnic University}, \orgaddress{\city{Hong Kong SAR}, \country{China}}}

\affil[16]{\orgdiv{Research Institute for Sports Science and Technology}, \orgname{The Hong Kong Polytechnic University}, \orgaddress{\city{Hong Kong SAR}, \country{China}}}


\abstract{
Alzheimer's disease (AD) dementia assessment in routine practice requires integrating heterogeneous, often incomplete clinical evidence while distinguishing AD from normal cognition (NC), mild cognitive impairment (MCI), and non-AD dementias. 
Existing multimodal models typically collapse this evidence into end-point
predictions, lacking the staged, criteria-based differential reasoning that
clinical assessment requires, whereas generic large language model (LLM)
reasoning remains clinically underconstrained and unreliable under
incomplete inputs. Here we present AD-CARE (\textbf{A}lzheimer's \textbf{D}isease \textbf{C}linical \textbf{A}ssessment and \textbf{Re}asoning Agent), a criteria-guided LLM agent that performs four-class differential diagnosis from heterogeneous routine clinical data. AD-CARE structures each case through a criteria-guided reasoning scaffold of evidence organization, stage-aware syndromic staging, and dementia-only etiologic arbitration, coordinating six diagnostic skills that span cognitive-functional, behavioral-neuropsychiatric, motor-neurological, vascular, MRI, and genetic evidence and operate on whichever modalities are present without imputation. 
Across 59,019 evaluable cases from five cohorts, on a four-class
differential task for which unassisted neurologists and radiologists in
our paired reader study achieved 50.0\% accuracy, AD-CARE reached
68.3\%--85.0\% accuracy, ranking first in all five cohorts and leading
macro-F1 in four of five. Gains were concentrated where differentiation
is hardest: relative to the baseline average, mean F1 was 7.1\% higher
for AD and 13.3\% higher for non-AD dementia, with non-AD
false-positive rates below 1\% in AD-focused cohorts. AD-CARE achieved
the highest accuracy in every one of 12 age, sex, race, and education
subgroups, and its worst subgroup matched or exceeded the best subgroup
of the strongest baselines. It retained a macro-F1 of 0.652 even in
high-missingness cases and lifted five standalone backbone LLMs by
2.29\%--10.66\%. With AD-CARE support, pooled clinician accuracy rose
by 19.3 percentage points (from 50.0\% to 69.3\%) and median reading
time fell 1.48-fold.
These results support criteria-guided LLM agents as a practical route toward scalable and equitable clinical decision support for AD-centered dementia assessment from routine real-world data.

}




\maketitle
\section{Introduction}\label{sec1}

Alzheimer's disease (AD) is the leading cause of dementia worldwide and a major contributor to progressive loss of memory, cognition, and functional independence~\cite{AD_lancet, joe2019cognitive, vos2017systematic}. As populations age, the number of affected individuals is projected to rise substantially, intensifying pressure on patients, families, and healthcare systems~\cite{mehta2017systematic, li2022global, marvi2024alzheimer}. Timely and accurate diagnosis is therefore clinically consequential, particularly as etiologic precision increasingly bears on therapeutic selection and downstream management. Amyloid and tau positron emission tomography (PET) and cerebrospinal fluid (CSF) biomarkers can increase biological confidence in AD diagnosis, but they do not remove the practical bottleneck of first-line dementia assessment. These tests remain costly, invasive or logistically constrained, are selectively ordered in routine care, and are not uniformly accessible across healthcare systems; blood-based biomarkers, although promising, are still being integrated into clinical workflows~\cite{AD_lancet, dubois2021clinical}. Accordingly, first-line dementia assessment is typically anchored in \emph{heterogeneous routine clinical data}: demographics, medical and vascular history, neuropsychological assessment, functional status, behavioral and neurological findings, and structural imaging, while biomarker and genetic tests may complement this assessment when available~\cite{mckhann2011diagnosis, dubois2021clinical}. These routine data are not merely a fallback when advanced testing is unavailable; they define syndrome stage, frame the initial differential diagnosis, and help determine whether confirmatory biomarker testing is warranted. This creates a practical need for decision-support systems that can operate directly on heterogeneous routine clinical data rather than assuming an idealized biomarker-complete setting.

Recent machine learning and deep learning methods have shown promise for AD-related prediction tasks, but their clinical usefulness remains limited in several important respects~\cite{aghdam2025machine, Christodoulou2025, odusami2024machine}. Most reported systems are unimodal, typically trained on a single source such as MRI, PET, or cognitive scores, and are frequently evaluated in highly curated research cohorts~\cite{aghdam2025machine, xun2025ada, bron2021cross}. Even when multimodal frameworks are used~\cite{Venugopalan2021, Qiu2022, yang2025adfound, Yuan2012, Aghili2022, Liu2023, yang2026taming, yang2026poe, yang2026brainanytime}, their dominant design remains predictive rather than decisional: heterogeneous evidence is fused primarily to optimize an end-point label, not organized into an explicit diagnostic sequence that separates syndrome-level staging from dementia-subtype attribution. Outputs likewise remain predominantly flat labels, scores, or probability vectors, offering limited visibility into how discordant evidence was reconciled and little explicit anchoring to disease-specific criteria for distinguishing AD from non-AD dementias~\cite{Martin2023, TaiyebKhosroshahi2025}. This limitation becomes most consequential precisely in the pre-biomarker or selectively biomarker-confirmed setting, where uncertainty is highest and diagnostic support must be extracted from incomplete routine evidence rather than from a uniform advanced testing panel. As a result, even accurate multimodal models may remain difficult to trust, audit, or integrate into real workflows for differential dementia assessment.

Large language models (LLMs) have introduced a complementary paradigm for clinical artificial intelligence because they can aggregate free-text information, follow complex instructions, invoke tools, and perform multi-step reasoning over heterogeneous inputs~\cite{Singhal2023, Maity2025, Liang2025}. This flexibility is especially attractive for dementia assessment in the selective-biomarker setting, where the diagnostic record often arrives as an opportunistic mixture of structured scores, narrative findings, routine imaging, and variable test availability rather than as a fixed biomarker panel. Yet generic LLM reasoning does not by itself solve differential dementia assessment. Without disease-specific constraints, generic LLMs can produce clinically plausible but criteria-inconsistent judgments, blur syndrome staging with etiologic attribution, and handle incomplete routine cases unreliably. Existing AD-related uses of LLMs therefore remain narrow in scope. One line of work applies large language models to clinical notes or speech to derive embeddings or textual markers for predicting early cognitive decline or progression, usually from a single modality and without engaging the full diagnostic work-up~\cite{Mao2023ADBERT, Guan2025CDTron, Mo2024LLMMarkers, Heitz2025GPT4Speech}. Another line develops AD-focused conversational systems, such as MemoryCompanion, AD-GPT, and AD-AutoGPT, that support question answering, information retrieval, or caregiving tasks rather than formal diagnostic assessment~\cite{Zheng2023MemoryCompanion, Liu2025ADGPT, Dai2025ADAutoGPT}. Recent reviews similarly note limited integration of imaging and biomarker evidence, sparse alignment with published diagnostic criteria, and minimal reader-study validation in routine memory-clinic settings~\cite{Treder2024LLMsDementia, Harrison2025LLMsDementia}. More recent agentic systems, including CARE-AD and ADAgent, demonstrate that LLMs can coordinate specialist roles or tool use for dementia-related tasks, but they still address only restricted slices of the diagnostic problem, such as longitudinal text-based prediction or narrow imaging-focused classification~\cite{Li2025CARE, hou2025adagent,zhao2026deeprare}Health. These advances establish feasibility, but they do not yet resolve the central challenge of real-world dementia assessment: how to organize incomplete multimodal evidence into a clinically structured differential diagnosis.

The central unresolved gap, therefore, is not multimodal prediction alone, nor generic reasoning capacity alone, but clinically explicit differential reasoning. In memory-clinic practice, dementia diagnosis follows a characteristic clinical structure. Clinicians first organize the available evidence, then determine the syndrome stage, and only thereafter perform etiologic differentiation among dementia subtypes when the case has crossed the dementia threshold. In other words, the key bottleneck is not simply whether a model can ``reason'', but whether the reasoning process is aligned with the staged logic of clinical decision-making. A system that sends every case directly to a flat end-point classifier, or that asks a generic LLM to infer the entire answer in one step, risks conflating syndrome-level staging with dementia-subtype attribution, over-calling rare alternatives, and failing to exploit the subset of evidence that is actually present. We therefore hypothesized that a more appropriate architecture for AD-centred decision support is a \emph{criteria-guided differential reasoning scaffold}: one that preserves the sequential logic of specialist assessment, uses stage-aware routing to separate syndromic staging from dementia-only arbitration, and grounds dementia differentiation in explicit consensus criteria rather than unconstrained free-form inference.

\textbf{Here, we present AD-CARE (Alzheimer's Disease Clinical Assessment and Reasoning Agent), a criteria-guided LLM agent framework for Alzheimer's disease dementia assessment from heterogeneous routine clinical data.} AD-CARE operationalizes this assessment as a four-class differential diagnosis problem (NC/MCI/AD/non-AD) through six structured diagnostic skills spanning cognitive-functional, behavioral-neuropsychiatric, motor-neurological, vascular, MRI-atrophy, and genetic risk evidence. These structured skills are coupled to a criteria-guided differential reasoning scaffold that performs cross-modal evidence organization, applies a stage-aware syndromic staging gate to distinguish NC, MCI, and Dementia, and invokes dementia-only specialist arbitration aligned with published disease-specific clinical criteria. Rather than assuming a fixed biomarker panel or complete work-up, AD-CARE conditions explicitly on the evidence that is available and omits unavailable modalities rather than imputing them. The system then returns a structured clinical output consisting of a four-class diagnosis (NC/MCI/AD/non-AD), a confidence estimate, and an evidence summary. In this formulation, the LLM serves not as a stand-alone classifier, but as the coordinating engine of a clinically constrained reasoning process.

We evaluated AD-CARE on 59,019 evaluable cases from five cohorts spanning public benchmarks and independent in-house clinical populations, with direct four-class testing in cohorts containing non-AD labels and specificity-oriented evaluation in AD-focused three-class cohorts. We further examined class-wise performance, equitable behavior across demographic subgroups, robustness to naturally occurring missingness, component-level ablations, portability across different backbone models, and potential clinical utility in a controlled reader study. Across these analyses, our aim was not merely to show incremental improvement on standard NC/MCI/AD classification, but to test whether a criteria-guided differential reasoning scaffold can address the hardest and most clinically relevant part of the task: differential dementia diagnosis under incomplete heterogeneous routine clinical evidence. Together, these results position AD-CARE as a clinically structured agent framework for AD-centered dementia assessment from heterogeneous routine clinical data.

\section{Results}\label{sec2}

\subsection{Overview of the AD-CARE framework}\label{system_overview}

In clinical practice, etiological assessment and staging of cognitive impairment require joint interpretation of heterogeneous routine clinical evidence, including demographics, functional assessments, neuropsychological testing, behavioral and neurological findings, structural neuroimaging, and genetic information. Dementia specialists typically (i) define the diagnostic question and available investigations, (ii) review each modality with domain-specific expertise, and (iii) reconcile concordant and discordant findings against established diagnostic criteria to arrive at an etiological diagnosis and disease stage.

AD-CARE operationalizes this process as a criteria-guided LLM agent framework for Alzheimer's disease dementia assessment from heterogeneous routine clinical data. Concretely, it implements this assessment as a four-class differential diagnosis problem (NC/MCI/AD/non-AD) under incomplete multimodal evidence. Rather than mapping raw inputs directly to a flat prediction, the framework centers on a \textbf{criteria-guided differential reasoning scaffold} coupled to \textbf{six diagnostic skills}. These six internal skills span cognitive-functional, behavioral-neuropsychiatric, motor-neurological, vascular risk, MRI-atrophy, and genetic risk reasoning. Each skill converts one aspect of the case into clinically interpretable evidence, such as cognitive severity signals, neuropsychiatric syndromes, parkinsonian and other neurological features, vascular burden, regional atrophy patterns, or genetic support for AD risk.

The criteria-guided differential reasoning scaffold serves as the organizing layer of the framework. It first performs \textbf{evidence organization}, then applies a \textbf{stage-aware syndromic staging gate} to distinguish NC, MCI, and Dementia, and finally performs \textbf{etiologic differential arbitration} for dementia cases. In this way, AD-CARE does not treat syndrome-level staging and etiologic differentiation as isolated tasks. Instead, the same scaffold selectively engages the relevant diagnostic skills, weighs concordant and discordant findings, and refines the final interpretation across candidate disease hypotheses. For dementia cases, this arbitration step is aligned with specialist reasoning for AD, LBD, FTD, and VaD, allowing etiologic differentiation to be grounded in explicit clinical criteria rather than generic free-form inference alone.

A defining property of AD-CARE is its flexible-input design for heterogeneous routine clinical data. The framework neither assumes a fixed input panel nor imputes unavailable modalities. Instead, the scaffold conditions explicitly on the subset of evidence that is present and engages the corresponding diagnostic skills as needed, allowing the same architecture to support syndrome-level staging, dementia differential diagnosis, and structured report generation under variable real-world data availability. The final output is a \textbf{structured clinical output} comprising the four-class diagnosis, a confidence estimate, and an evidence summary that highlights the key supporting and opposing findings. This design is clinically relevant because real-world workups are heterogeneous and often incomplete, yet the distinction between AD and non-AD dementia increasingly bears directly on treatment selection and downstream management.

\begin{figure}[!ht]
\centering
\includegraphics[width=1\textwidth]{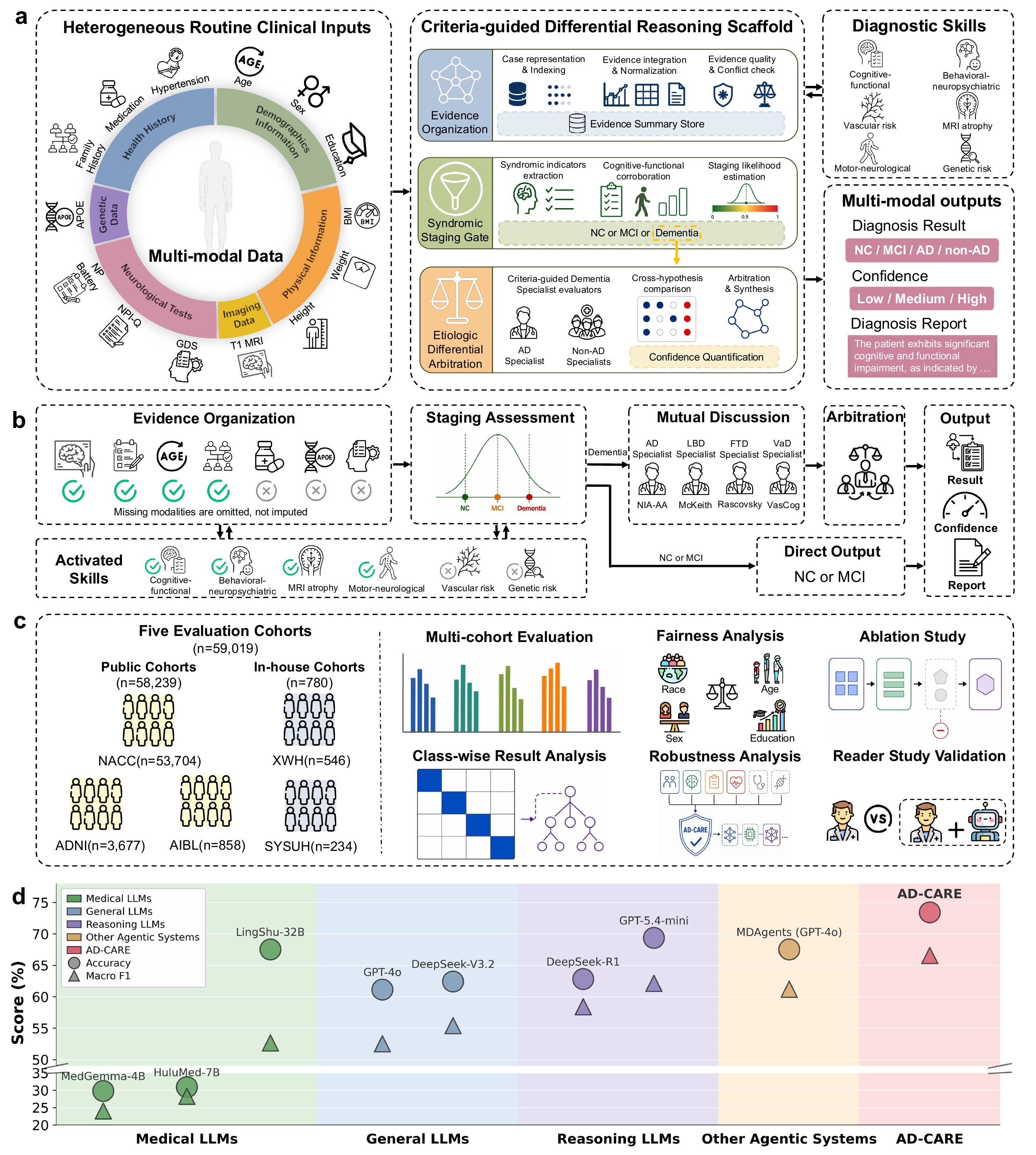}
\caption{\textbf{AD-CARE overview, evaluation program, and benchmark positioning.} \textbf{(a)}, Conceptual architecture of AD-CARE. Heterogeneous routine clinical inputs are organized by a criteria-guided differential reasoning scaffold with three functional layers: evidence organization, stage-aware syndromic staging, and etiologic differential arbitration. The scaffold interacts with six diagnostic skills and produces structured multimodal outputs, including a four-class diagnosis, confidence estimate, and report. \textbf{(b)}, Case-level routing under incomplete evidence. Available modalities are retained, unavailable modalities are omitted rather than imputed, only the compatible diagnostic skills are activated, NC/MCI cases can exit directly after staging, and dementia cases proceed to specialist discussion and arbitration before final output. \textbf{(c)}, Evaluation program across five cohorts, together with the downstream analyses used in this study, including multi-cohort performance assessment, class-wise analysis, fairness analysis, robustness analysis, ablation study, and reader-study validation. \textbf{(d)}, Overall benchmark positioning of AD-CARE relative to medical LLMs, general LLMs, reasoning LLMs, and other agentic systems, shown using accuracy and macro-F1.}\label{overview}
\end{figure}

\subsection{Study cohorts and evaluation strategy}\label{study_population}

\begin{table}[ht]
\caption{Study population and demographic characteristics (n = 59,019).}\label{demographic_table}
\centering
\begin{tabular}{lllll}
\toprule
\textbf{Cohort (group)}& \textbf{\makecell[l]{Age (y),\\mean $\pm$ s.d.}}& \textbf{Male, n (\%)}& \textbf{\makecell[l]{Education (y),\\mean $\pm$ s.d.}} &\textbf{CDR, mean $\pm$ s.d.}\\
\midrule
\rowcolor{gray!20}
\multicolumn{5}{c}{\textit{Public cohorts}}\\
\multicolumn{5}{l}{\textbf{NACC [n=53,704]}} \\
NC [n = 23,564] & $69.36 \pm 10.62$ & 8,192, 34.76\% & $15.84 \pm 2.97$ & $0.05 \pm 0.16$ \\
MCI [n = 13,028] & $72.26 \pm 9.15$ & 6,420, 49.28\% & $15.28 \pm 3.39$ & $0.45 \pm 0.18$ \\
AD [n = 12,826] & $73.76 \pm 10.02$ & 5,632, 43.91\% & $14.30 \pm 3.84$ & $1.14 \pm 0.71$ \\
non-AD [n = 4,286] & $68.01 \pm 9.90$ & 2,591, 60.45\% & $15.13 \pm 3.36$ & $1.17 \pm 0.81$ \\
$P$ value  & $<1.0 \times 10^{-3}$ & $<1.0 \times 10^{-3}$ & $<1.0 \times 10^{-3}$ & $<1.0 \times 10^{-3}$ \\
\multicolumn{5}{l}{\textbf{ADNI [n=3,677]}} \\
NC [n = 1,576] & $70.57 \pm 7.21$ & 603, 38.26\% & $16.49 \pm 2.52$ & $0.01 \pm 0.08$ \\
MCI [n = 1,563] & $72.62 \pm 7.74$ & 876, 56.05\% & $16.00 \pm 2.71$ & $0.49 \pm 0.09$ \\
AD [n = 538] & $74.89 \pm 7.92$ & 301, 55.95\% & $15.23 \pm 2.93$ & $0.77 \pm 0.29$ \\
$P$ value  & $<1.0 \times 10^{-3}$ & $<1.0 \times 10^{-3}$ & $<1.0 \times 10^{-3}$ & $<1.0 \times 10^{-3}$ \\
\multicolumn{5}{l}{\textbf{AIBL [n=858]}} \\
NC [n = 609] & $72.78 \pm 6.29$ & 268, 44.01\% & N/A & $0.03 \pm 0.12$ \\
MCI [n = 144] & $75.16 \pm 6.88$ & 79, 54.86\% & N/A & $0.47 \pm 0.14$ \\
AD [n = 105] & $74.26 \pm 7.96$ & 43, 40.95\% & N/A & $0.95 \pm 0.51$ \\
$P$ value  & $<1.0 \times 10^{-3}$ & $3.9 \times 10^{-2}$ & N/A & $<1.0 \times 10^{-3}$ \\
\rowcolor{gray!20}
\multicolumn{5}{c}{\textit{In-house cohorts}}\\
\multicolumn{5}{l}{\textbf{XWH [n=546]}} \\
NC [n = 85] & $61.81 \pm 10.71$ & 32, 37.65\% & N/A & $0.00 \pm 0.00$ \\
MCI [n = 114] & $64.77 \pm 8.35$ & 46, 40.35\% & N/A & $0.44 \pm 0.17$ \\
AD [n = 270] & $64.33 \pm 8.33$ & 96, 35.56\% & N/A & $1.24 \pm 0.69$ \\
non-AD [n = 77] & $61.91 \pm 5.65$ & 43, 55.84\% & N/A & $1.26 \pm 0.84$ \\
$P$ value  & $1.9 \times 10^{-2}$ & $1.5 \times 10^{-2}$ & N/A & $<1.0 \times 10^{-3}$ \\
\multicolumn{5}{l}{\textbf{SYSUH [n=234]}} \\
NC [n = 20] & $61.85 \pm 12.49$ & 10, 50.00\% & $12.95 \pm 3.41$ & N/A \\
MCI [n = 31] & $67.00 \pm 9.51$ & 17, 54.84\% & $11.97 \pm 4.20$ & N/A \\
AD [n = 165] & $67.40 \pm 10.40$ & 65, 39.39\% & $9.04 \pm 5.15$ & N/A \\
non-AD [n = 18] & $67.89 \pm 9.72$ & 9, 50.00\% & $7.39 \pm 4.96$ & N/A \\
$P$ value  & $2.9 \times 10^{-1}$ & $3.3 \times 10^{-1}$ & $<1.0 \times 10^{-3}$ & N/A \\
\botrule
\end{tabular}
\end{table}

We evaluated AD-CARE on 59,019 evaluable cases from five independent cohorts (Table~\ref{demographic_table}): NACC~\cite{NACC}, ADNI~\cite{ADNI}, AIBL~\cite{AIBL}, and two in-house clinical cohorts from Xuanwu Hospital (XWH) and the Third Affiliated Hospital of Sun Yat-sen University (SYSUH). Ground-truth non-AD labels were available in NACC (n = 4,286), XWH (n = 77), and SYSUH (n = 18), enabling direct evaluation of non-AD detection sensitivity.

A central design principle of this evaluation is that AD-CARE always emits four-class predictions regardless of the cohort label space. In the four-class cohorts (NACC, XWH, SYSUH), we therefore assessed direct NC/MCI/AD/non-AD differential diagnosis. In the three-class cohorts (ADNI, AIBL), we evaluated conventional NC/MCI/AD classification while treating any non-AD prediction as a false positive. These two settings jointly test whether AD-CARE can both detect non-AD dementia when it is present and avoid spuriously over-calling it when it is absent.

We evaluated AD-CARE (GPT-4o backbone) against a broader baseline panel comprising three medical LLMs (MedGemma-4B, HuluMed-7B, and LingShu-32B), two general-purpose LLMs (DeepSeek-V3.2 and GPT-4o), two reasoning-oriented models (DeepSeek-R1 and GPT-5.4-mini), and a general-purpose medical agent framework (MDAgents with GPT-4o backbone). For clarity in the main-text cohort-level analyses below, we then focused on the five strongest representative baselines: DeepSeek-V3.2, GPT-4o, DeepSeek-R1, GPT-5.4-mini, and MDAgents. Complete results for all methods are provided in Supplementary Table~4.

\subsection{Multi-Cohort Diagnostic Assessment on Public Cohorts}\label{main_results}

\begin{figure}[!ht]
\centering
\includegraphics[width=1\textwidth]{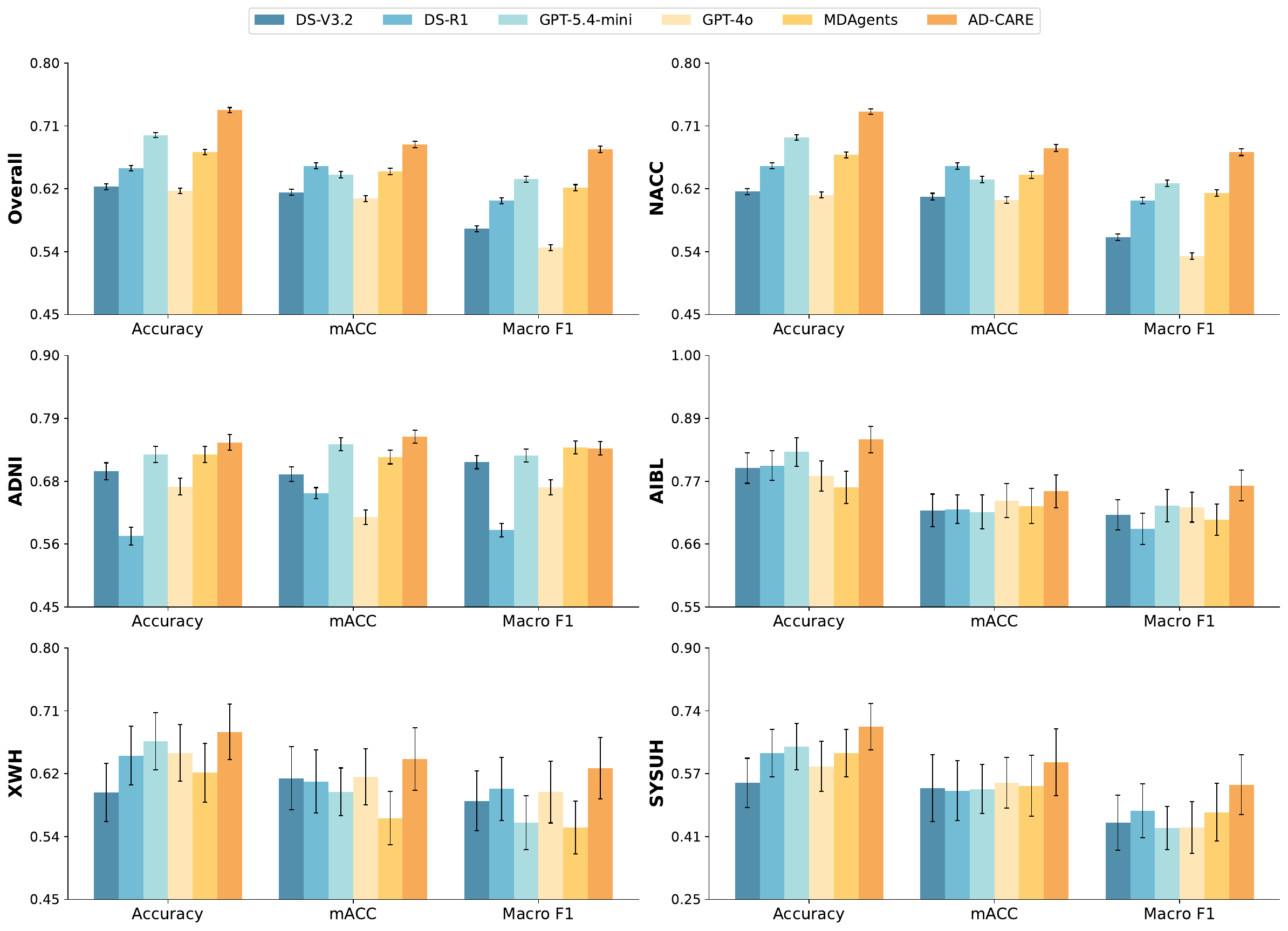}
\caption{\textbf{Overall diagnostic performance of AD-CARE and baseline methods across five cohorts.} Each panel reports accuracy, mean class accuracy (mACC), and macro-F1 for six methods. Error bars denote 95\% bootstrap confidence intervals (B = 2,000). AD-CARE achieves the highest accuracy in all five cohorts and the highest macro-F1 in four of five; on ADNI, macro-F1 is essentially tied with MDAgents.}\label{overall_performance}
\end{figure}

To demonstrate the effectiveness of AD-CARE, we performed a multi-cohort evaluation on three public cohorts in Fig.~\ref{overall_performance}. The complete results for all methods are provided in Supplementary Table~1. \textbf{Across these public benchmarks, AD-CARE was the only method that remained first in both accuracy and mean class accuracy (mACC) in every cohort, while also delivering the highest macro-F1 in NACC and AIBL and near-parity with the top macro-F1 on ADNI.}

The clearest separation emerged on NACC, the largest and most diagnostically heterogeneous public cohort. Here, AD-CARE achieved an accuracy of 0.733 [95\% CI: 0.729, 0.736], an mACC of 0.682 [95\% CI: 0.677, 0.686], and a macro-F1 of 0.676 [95\% CI: 0.672, 0.680], whereas the next-best overall method, GPT-5.4-mini, reached 0.697 [95\% CI: 0.693, 0.701], 0.638 [95\% CI: 0.633, 0.643], and 0.633 [95\% CI: 0.628, 0.638], respectively. Thus, the advantage on NACC was visible simultaneously in overall correctness, class-balanced recall, and balanced precision-recall performance, indicating that AD-CARE is particularly effective in more heterogeneous and clinically demanding differential diagnosis settings.

AD-CARE remained similarly competitive on the two public three-class cohorts. On ADNI, it ranked first in accuracy and mACC, reaching 0.744 [95\% CI: 0.731, 0.759] and 0.755 [95\% CI: 0.740, 0.770], while its macro-F1 of 0.734 [95\% CI: 0.719, 0.750] was essentially tied with MDAgents (0.736 [95\% CI: 0.720, 0.751]). This pattern is important because it shows that extending the diagnostic space to include non-AD did not compromise standard NC/MCI/AD discrimination. On AIBL, AD-CARE regained a clear lead across all three metrics, achieving an accuracy of 0.850 [95\% CI: 0.825, 0.873], an mACC of 0.757 [95\% CI: 0.717, 0.796], and a macro-F1 of 0.767 [95\% CI: 0.730, 0.803], compared with 0.828 [95\% CI: 0.802, 0.852], 0.720 [95\% CI: 0.680, 0.758], and 0.731 [95\% CI: 0.694, 0.766] for GPT-5.4-mini. \textbf{Taken together, these results indicate that AD-CARE preserves excellent performance on conventional AD-focused cohorts while showing its largest gains on the more heterogeneous four-class benchmark, precisely where differential diagnosis is most demanding.}

\subsection{Multi-Cohort Diagnostic Assessment on In-house Cohorts}\label{main_results2}

We next evaluated AD-CARE on two clinically collected in-house cohorts, XWH and SYSUH, to probe generalization under real-world data heterogeneity. As shown in Fig.~\ref{overall_performance}, baseline methods exhibited visibly wider uncertainty on these cohorts, consistent with the increased variability typically encountered in routine clinical settings. \textbf{Despite this shift, AD-CARE remained the top-performing method in both cohorts, ranking first in accuracy, mean class accuracy (mACC), and macro-F1 in each cohort.}

In XWH, AD-CARE achieved an accuracy of 0.683 [95\% CI: 0.645, 0.722], an mACC of 0.645 [95\% CI: 0.602, 0.686], and a macro-F1 of 0.633 [95\% CI: 0.589, 0.671], again delivering the strongest overall profile among all compared methods. In SYSUH, AD-CARE achieved an accuracy of 0.697 [95\% CI: 0.633, 0.756], an mACC of 0.604 [95\% CI: 0.519, 0.686], and a macro-F1 of 0.546 [95\% CI: 0.469, 0.619], likewise ranking first across all three metrics. Notably, the separation remained at least as clear on SYSUH, where uncertainty widened further for the baselines, suggesting that the benefit of structured, criteria-guided reasoning reasoning becomes more apparent as real-world heterogeneity and diagnostic ambiguity increase.

\textbf{These results indicate that AD-CARE maintains a high and balanced diagnostic performance even in independent clinical populations collected outside public benchmark settings, supporting its translational relevance for real-world deployment.}

\begin{figure}[!htbp]
\centering
\includegraphics[width=1\textwidth]{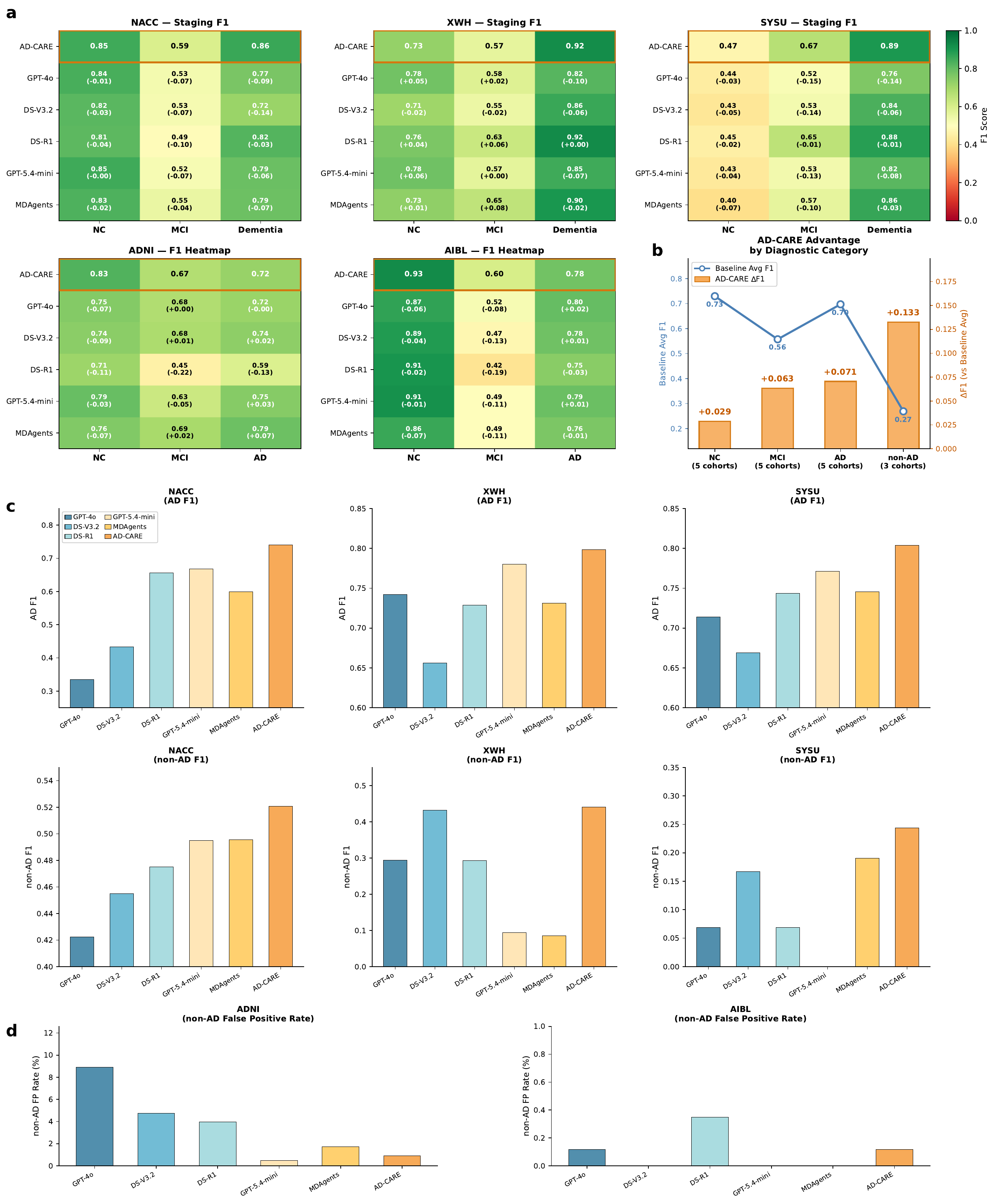}

\caption{\textbf{Class-wise diagnostic analysis shows that AD-CARE's advantage is concentrated in dementia differentiation.} \textbf{(a)}, Per-class F1 heatmaps for the four-class cohorts (NACC, XWH, SYSUH) under the three-class staging view (NC / MCI / Dementia; AD and non-AD merged). Rows correspond to methods and columns to diagnostic categories; separation across methods is smaller at this syndrome level than in the etiologic comparisons below. \textbf{(b)}, Per-class F1 heatmaps for the three-class cohorts (ADNI, AIBL) and AD-CARE advantage by diagnostic category (right), showing a monotonic increase from NC (+0.029) to non-AD (+0.133). \textbf{(c)}, AD and non-AD F1 across the four-class cohorts. \textbf{(d)}, Non-AD false-positive rates on ADNI and AIBL; AD-CARE remains below 1\% on both.}\label{diagnostic_detail}
\end{figure}

\subsection{Class-wise Diagnostic Analysis}\label{nonad_analysis}
Overall cohort-level metrics do not reveal which diagnostic categories account for the observed improvement. We therefore performed a class-wise analysis across all five cohorts to identify which categories contributed most to AD-CARE's performance gains (Fig.~\ref{diagnostic_detail}). To separate syndrome-level staging from etiologic differentiation, we first evaluated the four-class cohorts under a three-class view (NC / MCI / Dementia, with AD and non-AD merged; Fig.~\ref{diagnostic_detail}a). At this staging level, differences between AD-CARE and the baselines were comparatively modest. On NACC, AD-CARE remained first, with a Dementia F1 of 0.856 versus 0.822 for DeepSeek-R1, while NC was essentially tied with GPT-5.4-mini (0.849 versus 0.848). On the in-house cohorts, staging performance was more mixed, and several baselines already achieved high Dementia F1 once etiologic labels were collapsed. The three-class cohorts (ADNI, AIBL) likewise showed smaller overall separation across methods. \textbf{Together, these heatmap-level comparisons indicate that AD-CARE's advantage is not explained by superior NC/MCI/Dementia staging, but emerges in the subsequent problem of dementia differentiation.}

This interpretation was confirmed when the gain was averaged by diagnostic category rather than by staging (Fig.~\ref{diagnostic_detail}b). The mean AD-CARE advantage over the baseline average increased from NC ($\Delta$F1 = +0.029) to MCI (+0.063) and AD (+0.071), and was largest for non-AD (+0.133). Thus, the harder the diagnostic category, the larger the benefit conferred by structured clinical reasoning. In particular, the larger gains in AD and non-AD relative to NC and MCI indicate that the principal advantage of AD-CARE lies not in generic classification uplift, but in more effective dementia differentiation.

Within this broader dementia-differentiation pattern, non-AD discrimination was the strongest single contributor (Fig.~\ref{diagnostic_detail}c). On NACC, AD-CARE achieved a non-AD F1 of 0.521, compared with 0.496 for MDAgents, 0.495 for GPT-5.4-mini, 0.475 for DeepSeek-R1, 0.455 for DeepSeek-V3.2, and 0.422 for GPT-4o. The same pattern persisted in the in-house cohorts. On XWH, AD-CARE achieved a non-AD F1 of 0.441, slightly above the next-best method (DeepSeek-V3.2, 0.432), whereas several baselines fell below 0.10. On SYSUH, AD-CARE reached a non-AD F1 of 0.244, compared with 0.191 for MDAgents, while GPT-5.4-mini failed to detect any non-AD case (F1 = 0.000). These results indicate that the largest absolute separation between AD-CARE and the baselines arises from recognizing dementias that do not follow the canonical AD pattern.

A reasonable concern is that higher non-AD sensitivity might simply reflect over-prediction of the non-AD label. To examine this possibility, we quantified the non-AD false-positive rate on ADNI and AIBL (Fig.~\ref{diagnostic_detail}d), where all ground-truth labels are NC, MCI, or AD. AD-CARE produced non-AD false-positive rates of 0.90\% on ADNI and 0.12\% on AIBL, whereas GPT-4o reached 8.92\% on ADNI. Taken together with the stronger class-wise performance in AD and non-AD, these comparatively low false-positive rates suggest that AD-CARE does not simply over-predict non-AD dementia, but maintains a relatively specific non-AD decision profile.

\begin{figure}[!ht]
\centering
\includegraphics[width=1\textwidth]{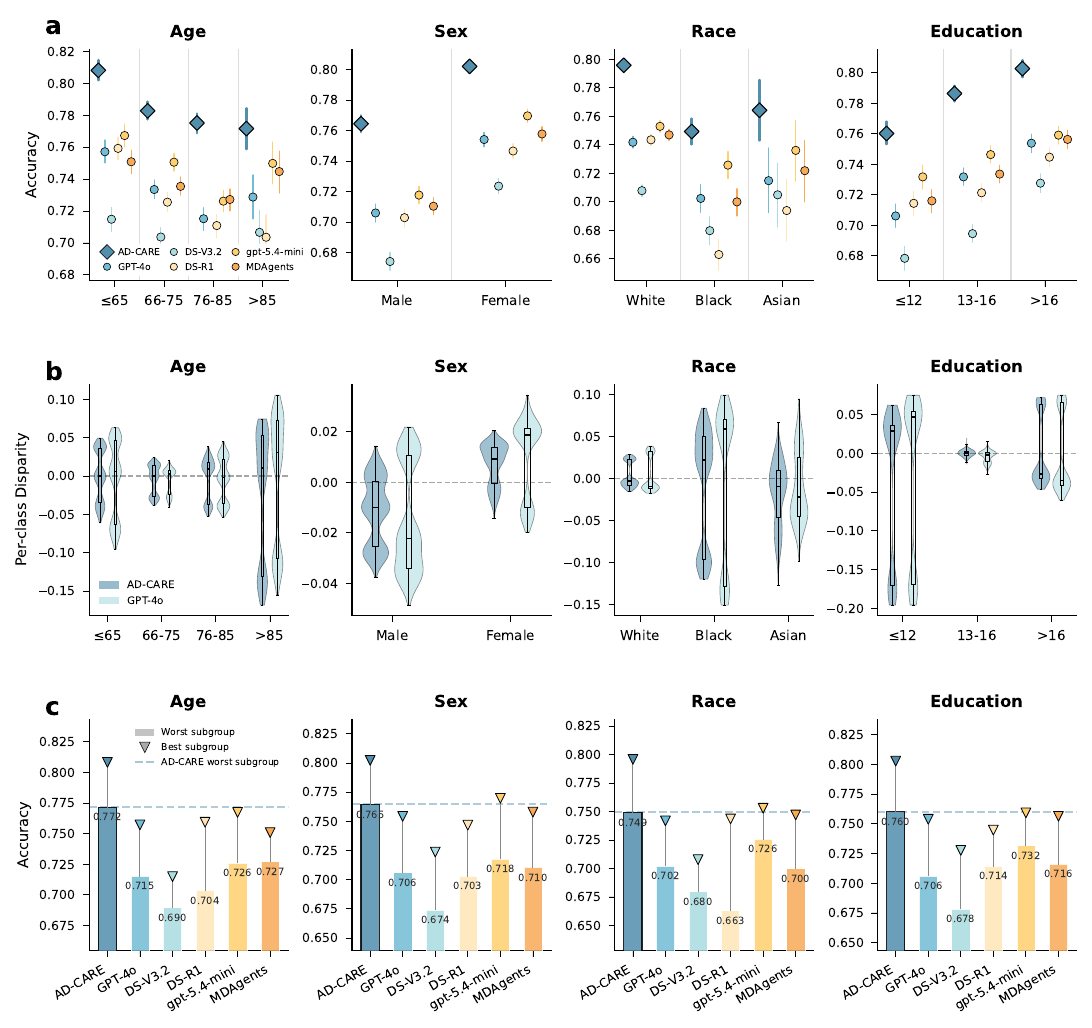}
\caption{\textbf{Equitable performance across demographic subgroups in pooled NACC and ADNI analyses.} All panels are evaluated in the shared three-class space (NC / MCI / Dementia). \textbf{(a)}, Per-subgroup accuracy with 95\% bootstrap confidence intervals across age, sex, race, and education. AD-CARE achieves the highest accuracy in all 12 pooled subgroups. \textbf{(b)}, Per-class disparity distributions (subgroup sensitivity minus overall sensitivity) for AD-CARE versus GPT-4o; narrower violins indicate more consistent class-specific detection across subgroups. \textbf{(c)}, Worst-subgroup accuracy (bars) and best-subgroup accuracy (triangles) for each method. AD-CARE maintains the strongest or near-strongest worst-subgroup performance across all four dimensions, with the clearest separation in age.}\label{fairness_ana}
\end{figure}

\subsection{Equitable Performance Across Demographic Subgroups}\label{fairness}

To determine whether the diagnostic gains of AD-CARE were equitably distributed across demographic groups rather than concentrated in majority populations, we pooled NACC and ADNI for subgroup analysis, as these were the two cohorts with sufficiently large sample sizes and structured annotations for age, sex, race, and education. This yielded a combined evaluation set of 57,372 cases, with all available cases retained within each demographic dimension. For this analysis, we adopted a shared three-class scheme (NC / MCI / Dementia, with AD and non-AD merged into Dementia). This choice enabled a harmonized evaluation space across the two cohorts and provided more stable subgroup-level estimates after demographic stratification, particularly because non-AD cases become sparse in several subgroups, with more details in Supplementary Table~7. We then examined three complementary aspects of subgroup behavior: per-subgroup accuracy with 95\% bootstrap confidence intervals (Fig.~\ref{fairness_ana}a), per-class disparity relative to overall sensitivity (Fig.~\ref{fairness_ana}b), and worst-subgroup versus best-subgroup accuracy across methods (Fig.~\ref{fairness_ana}c).

Fig.~\ref{fairness_ana}a first shows that AD-CARE achieved the highest accuracy in all 12 pooled subgroups, indicating that its performance gain was broadly shared rather than limited to majority groups. This pattern extended to underrepresented subgroups, including Asian participants (n = 1,650; accuracy = 0.764 [95\% CI: 0.743, 0.785]) and adults older than 85 years (n = 4,142; accuracy = 0.772 [95\% CI: 0.759, 0.784]), supporting the view that the benefit of AD-CARE is not confined to easier demographic strata.

This advantage was complemented by the class-level disparity analysis in Fig.~\ref{fairness_ana}b. Compared with GPT-4o, AD-CARE showed tighter disparity distributions across age, sex, race, and education, indicating that subgroup-to-overall deviations in class-specific sensitivity were smaller and more consistent. Thus, the subgroup gains observed in Fig.~\ref{fairness_ana}a do not appear to arise from improvement in only one or two diagnostic classes, but instead reflect more stable detection behavior across categories.

Fig.~\ref{fairness_ana}c provides a stricter robustness summary by focusing on worst-subgroup accuracy. The clearest separation was observed for age: AD-CARE's worst-performing age subgroup ($>$85 years; accuracy = 0.772 [95\% CI: 0.759, 0.784]) still exceeded the strongest age subgroup achieved by any baseline (GPT-5.4-mini, $\leq$65 years; accuracy = 0.767 [95\% CI: 0.760, 0.774]). Across sex, race, and education, worst-subgroup accuracy also remained high: 0.765 [95\% CI: 0.759, 0.770] for males, 0.749 [95\% CI: 0.740, 0.758] for Black participants, and 0.760 [95\% CI: 0.753, 0.768] for participants with $\leq$12 years of education. In these dimensions, AD-CARE's worst subgroup was at or close to the strongest subgroup performance reached by the best baselines (sex, 0.770 [95\% CI: 0.765, 0.774]; race, 0.753 [95\% CI: 0.749, 0.757]; education, 0.759 [95\% CI: 0.753, 0.765]). \textbf{Taken together, these results indicate that AD-CARE improves diagnostic performance without concentrating that benefit in only a few demographic groups, supporting a pattern of equitable performance across clinically relevant population strata.}

\subsection{Robustness and Component Contribution Analyses}\label{robustness}
\begin{figure}[!ht]
\centering
\includegraphics[width=1\textwidth]{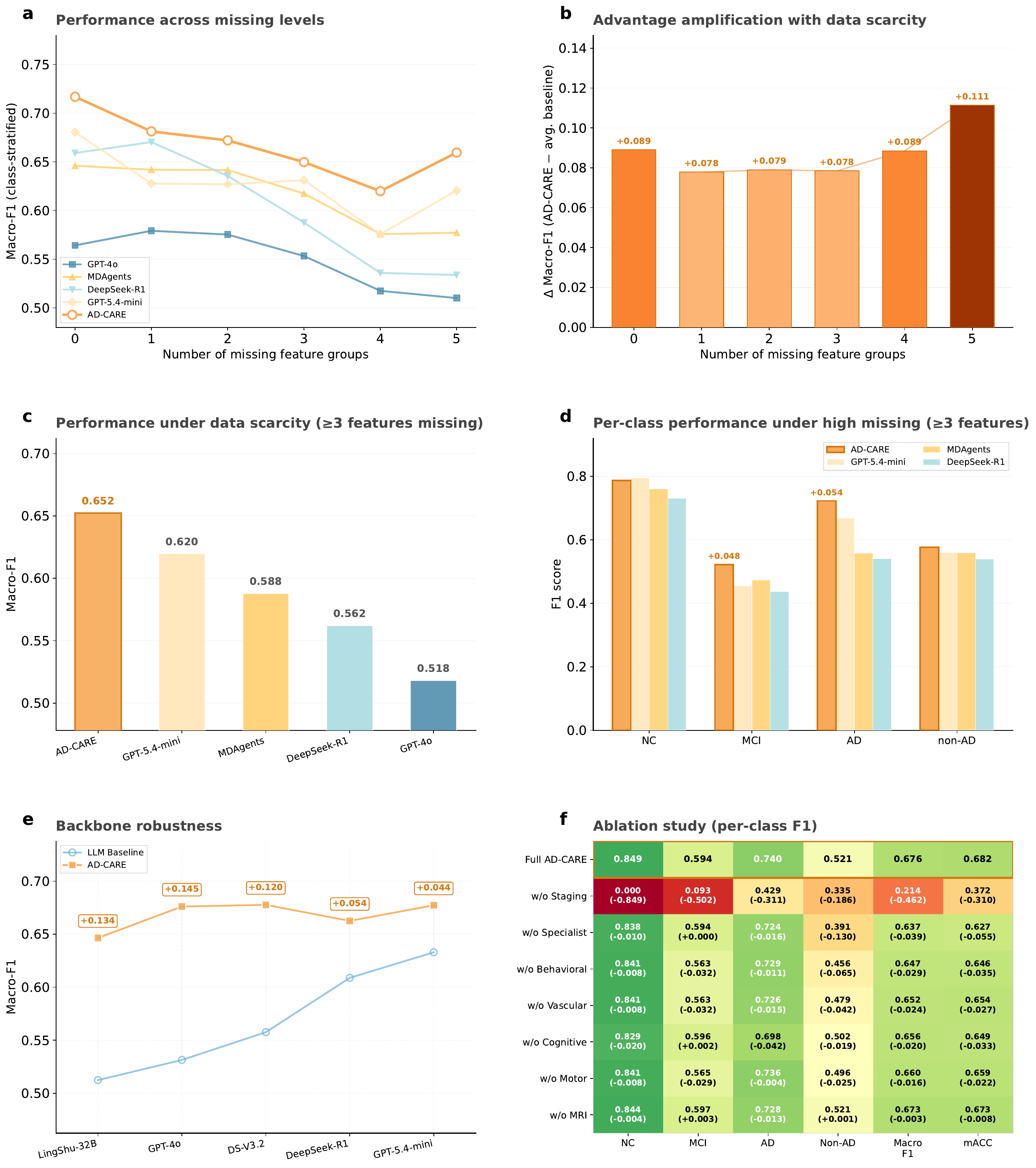}
\caption{\textbf{Robustness and component contribution analyses of AD-CARE.} \textbf{(a)}, Macro-F1 across missing feature levels for five methods. \textbf{(b)}, AD-CARE's advantage over the baseline average widens under severe missingness and peaks at +0.111 with five missing feature groups. \textbf{(c)}, Performance under high missingness ($\geq$3 missing feature groups). \textbf{(d)}, Per-class F1 under high missingness. \textbf{(e)}, Paired comparison of standalone LLM backbones versus the corresponding AD-CARE-augmented systems, showing consistent gains after agentic augmentation. \textbf{(f)}, Ablation heatmap showing per-class F1 and overall metrics after removing individual components relative to full AD-CARE.}\label{robustness_fig}
\end{figure}
To assess the deployment robustness of AD-CARE and identify the components underlying its performance gains, we evaluated the framework along three complementary axes in the order shown in Fig.~\ref{robustness_fig}: naturally occurring missingness (Fig.~\ref{robustness_fig}a--d), backbone variation (Fig.~\ref{robustness_fig}e), and component ablation (Fig.~\ref{robustness_fig}f). \textbf{Together, these analyses distinguish deployment robustness from internal component contribution and show that AD-CARE's diagnostic advantage is preserved across incomplete inputs and changing model backbones.}

\textbf{\textit{Missing data.}} Fig.~\ref{robustness_fig}a,b first examine naturally occurring missingness in NACC by stratifying cases according to the number of missing feature groups. AD-CARE retained the highest macro-F1 at every missingness level, and its advantage over the baseline average ranged from +0.078 to +0.111, peaking at the most severe level with five missing feature groups. Thus, the relative benefit of the framework became more pronounced, rather than attenuated, as the clinical record became sparser. Fig.~\ref{robustness_fig}c,d then focus on high-missingness cases ($\geq$3 missing feature groups). Even in this setting, AD-CARE achieved a macro-F1 of 0.652, compared with 0.620 for GPT-5.4-mini and 0.588 for MDAgents. The class-wise results in Fig.~\ref{robustness_fig}d further show that this advantage was not confined to a single category, but remained distributed across the diagnostic task. These findings indicate that AD-CARE degrades comparatively gracefully when substantial parts of the clinical record are absent.

\textbf{\textit{Backbone variation.}} Fig.~\ref{robustness_fig}e next asks whether the performance gain depends on a specific foundation model. Across all paired backbones included in this analysis, wrapping the same model with AD-CARE improved diagnostic performance relative to the corresponding standalone LLM. Although the magnitude of improvement varied across backbones, the direction of effect was uniformly positive, indicating that the agentic clinical scaffold contributes benefits that are not tied to a single model family. This result supports the practical portability of the framework across deployment settings with different cost, access, or governance constraints.

\textbf{\textit{Component ablation.}} Fig.~\ref{robustness_fig}f finally examines which components are responsible for these gains. Every component removal reduced overall macro-F1, but the magnitudes differed substantially. At the framework level, removing the staging module produced by far the largest collapse ($\Delta$macro-F1 = $-$0.462), indicating that explicit disease-stage organization is a central function of the framework rather than a superficial prompt layer. Removing the specialist layer caused the next-largest decline ($\Delta$macro-F1 = $-$0.039) and was associated with a marked reduction in non-AD F1 ($\Delta$ = $-$0.130), indicating that specialist-mediated arbitration contributes importantly to dementia differential diagnosis after the initial staging step. At the tool level, the behavioral ($-$0.029) and vascular ($-$0.024) modules provided the largest additional gains. By contrast, removing MRI caused only a small overall decrease ($\Delta$macro-F1 = $-$0.003), although the drop was more evident for AD-related recognition, consistent with the value of structural atrophy information. Together, these ablations show that AD-CARE's performance arises from coordinated contributions across modules, with staging providing the dominant organizing role and specialist reasoning making an additional framework-level contribution to dementia discrimination.

\subsection{AD-CARE enhances clinician diagnostic accuracy and efficiency}\label{reader_study}

To test whether these algorithmic gains translated into practical clinical benefit, we conducted a paired reader study with seven clinicians from Xuanwu Hospital, Capital Medical University (XWH) and the Third Affiliated Hospital of Sun Yat-sen University (SYSUH). Three readers were from XWH and four from SYSUH. Each independently reviewed 40 NACC cases in two sessions separated by a washout of 7 to 20 days, as described in Section~\ref{expert-level validation}. The set comprised ten cases each of NC, MCI, AD, and non-AD dementia, and 20 of the 40 cases included MRI. Readers first diagnosed from routine clinical data alone and, after washout, re-read the same cases together with AD-CARE's structured report. Cases were sampled to over-represent instances in which AD-CARE outperformed its GPT-4o backbone, so that the study would test whether that algorithmic advantage could be taken up in human decisions rather than reproduce clinic prevalence. \textbf{AD-CARE assistance improved pooled diagnostic accuracy and shortened reading time, with the largest time savings on the cases that readers otherwise disagreed about.}

\begin{figure}[!htbp]
\centering
\includegraphics[width=1\textwidth]{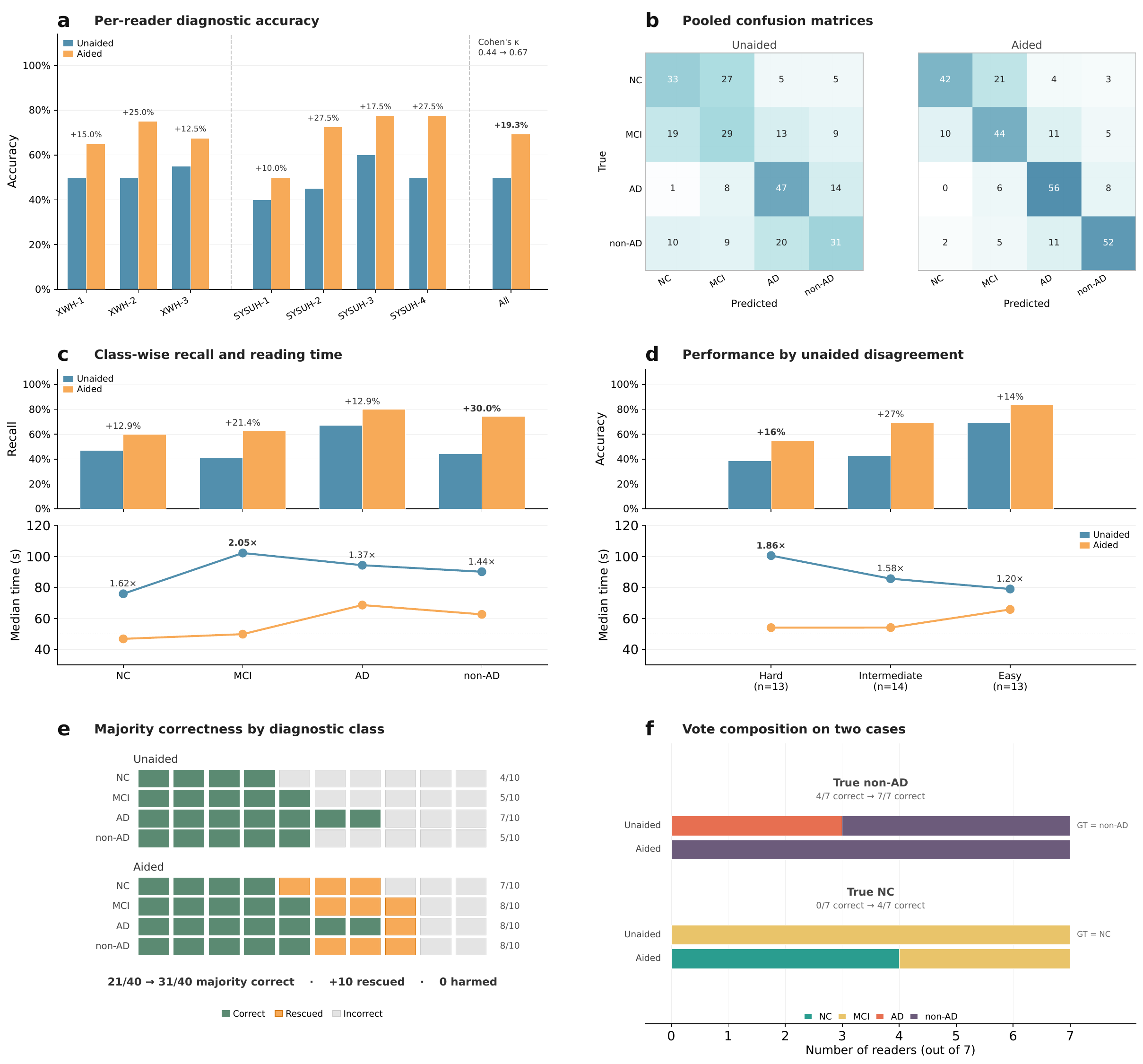}
\caption{\textbf{AD-CARE assistance improves clinician accuracy, speed, and panel consensus.} Steel blue denotes unaided reads and amber denotes assisted reads. Standalone AD-CARE accuracy on this enriched 40-case set was 80\% and is not used as a visual comparator. \textbf{(a)}, Per-reader diagnostic accuracy grouped by site, with pooled accuracy at right. Accuracy increased from 50.0\% to 69.3\% and pairwise Cohen's $\kappa$ from 0.44 to 0.67. All seven readers improved. \textbf{(b)}, Pooled confusion matrices of true versus predicted labels across 280 reads. Off-diagonal mass decreased most for AD versus non-AD and for MCI versus NC. \textbf{(c)}, Class-wise recall and median reading time. The largest recall gain was in non-AD, at 30.0 percentage points; MCI reads accelerated most, by 2.05-fold. \textbf{(d)}, Accuracy and median time after stratification by unaided inter-reader disagreement rather than by correctness. Accuracy rose in every tertile; Hard cases accelerated most, by 1.86-fold, and aided times converged near 50--60~s. \textbf{(e)}, Majority-vote correctness by diagnostic class, with ten cases per row. Unaided reads are shown above assisted reads. Majority-correct cases sit left, subsequently corrected cases in the middle, and still-incorrect cases at right. The seven-reader majority was correct for 21 of 40 cases unaided and 31 of 40 with assistance; ten cases were corrected and none were reversed. \textbf{(f)}, Vote composition for two illustrative cases. Right-hand labels give the ground-truth class. A true non-AD case collapsed to unanimous agreement, whereas a false unanimous MCI consensus on a true NC case was broken.}\label{user_study}
\end{figure}

Across 280 paired reads, pooled four-class accuracy increased from 50.0\% to 69.3\%, a gain of 19.3 percentage points, and macro-F1 rose from 0.497 to 0.693 (Fig.~\ref{user_study}a). An exact McNemar test showed that corrections outnumbered regressions, 69 versus 15, $P = 1.9 \times 10^{-9}$. Pairwise Cohen's $\kappa$ increased from 0.44 to 0.67. All seven readers improved, with individual gains ranging from 10.0 to 27.5 percentage points. Six of the seven were also faster. One reader was slightly slower, with median time rising from 108~s to 116~s, yet remained more accurate by 12.5 percentage points. Site-level gains were concordant, at 17.5 percentage points at XWH and 20.6 percentage points at SYSUH. Median reading time across all reads fell from 85.7~s to 57.9~s, a 1.48-fold reduction.

These gains were concentrated in etiologic differentiation and staging rather than in further improving recognition of NC (Fig.~\ref{user_study}c). Recall increased by 12.9, 21.4, 12.9, and 30.0 percentage points for NC, MCI, AD, and non-AD, respectively, taking non-AD recall from 44.3\% to 74.3\%. Median reading time fell most for MCI, by 2.05-fold. Pooled confusion matrices showed the same pattern, with reduced off-diagonal mass for AD versus non-AD and for MCI versus NC (Fig.~\ref{user_study}b). Reader--reader disagreements of AD versus non-AD fell from 106 pairs to 27, and those of MCI versus NC from 99 to 64. Accuracy improved both when MRI was available, by 20.0 percentage points, and when it was not, by 18.6 percentage points.

We next stratified cases by unaided inter-reader pairwise agreement, a measure of clinical controversy that correlated only moderately with unaided case-level accuracy; Spearman $\rho = 0.42$. Accuracy increased in every tertile: from 38\% to 55\% on Hard cases, from 43\% to 69\% on Intermediate cases, and from 69\% to 84\% on Easy cases (Fig.~\ref{user_study}d). Pairwise agreement rose from 32\% to 66\% on Hard cases and fell from 91\% to 86\% on Easy cases, where a population-level false consensus was broken. Assistance saved the most time where disagreement had been greatest. Median time on Hard cases fell from 101~s to 54~s, a 1.86-fold reduction, compared with 1.58-fold and 1.20-fold reductions on Intermediate and Easy cases. Unaided readers were already slower on high-disagreement than on low-disagreement cases, 101~s versus 79~s; with assistance, times converged near 50--60~s. The time saving was therefore realized on the records that clinicians otherwise contested, rather than on cases that were already straightforward.

At the case level, the seven-reader modal diagnosis was correct for 21 of 40 cases unaided and for 31 of 40 with assistance (Fig.~\ref{user_study}e). Ten cases moved from an incorrect to a correct majority, and none reversed from a correct to an incorrect majority. These net gains were distributed across classes, from 4 to 7 for NC, 5 to 8 for MCI, 7 to 8 for AD, and 5 to 8 for non-AD. Pairwise percent agreement rose from 58\% to 75\%. Assistance both consolidated true consensus and broke false consensus (Fig.~\ref{user_study}f). In one true non-AD case, the correct vote increased from four of seven readers to all seven. In one true NC case that all seven readers had labelled MCI, four switched to NC after assistance.

\begin{figure}[!t]
\centering
\includegraphics[width=1\textwidth]{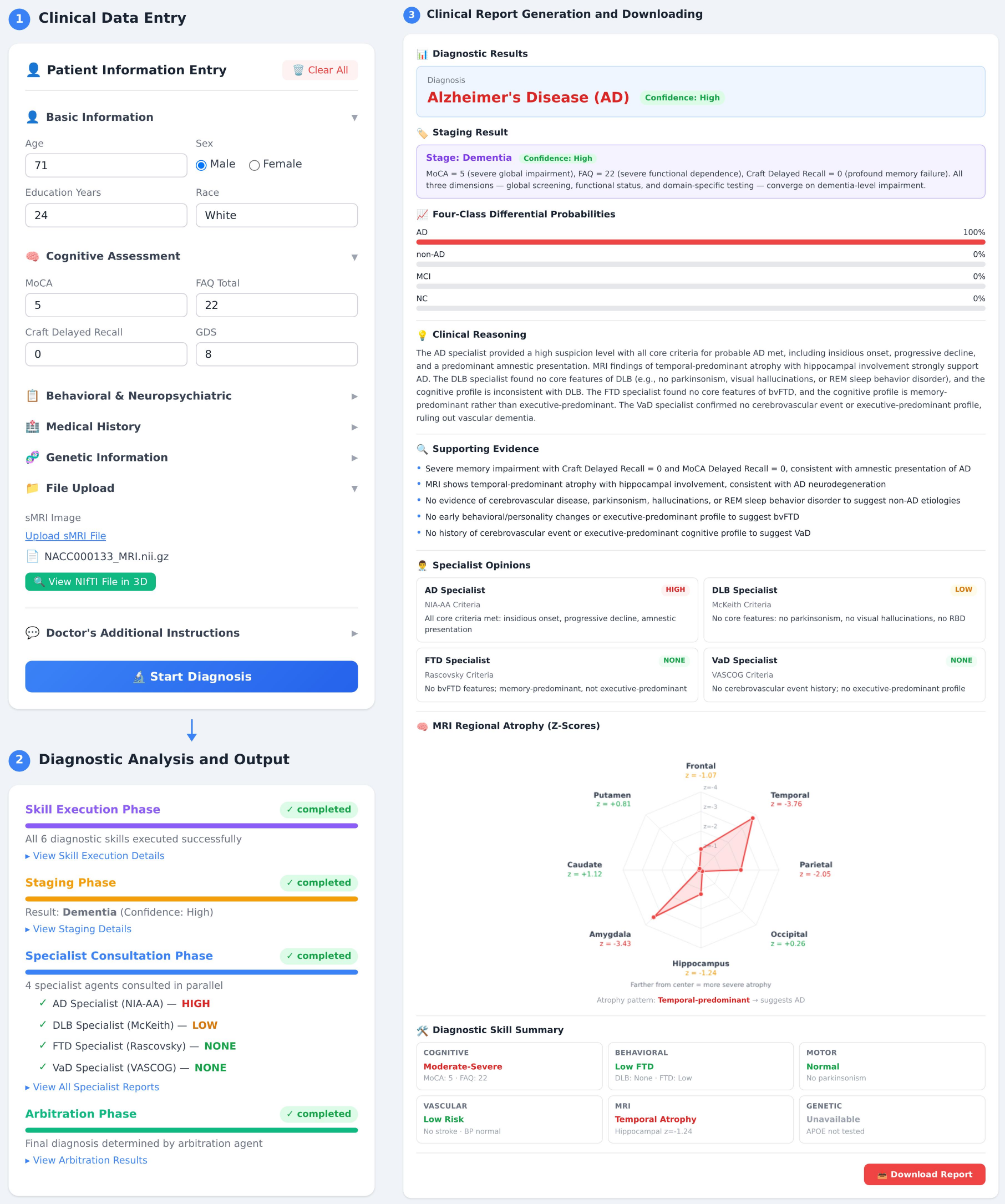}
\caption{\textbf{Web interface for clinical deployment of AD-CARE.} \textbf{(1)}, Data entry: users input demographics, neuropsychological scores, behavioral and neuropsychiatric assessments, medical history, APOE status, and functional assessments, with optional upload of structural MRI (NIfTI). \textbf{(2)}, Diagnostic analysis: the agent executes a four-phase pipeline --- skill execution $\to$ staging $\to$ specialist consultation $\to$ arbitration --- with transparent progress tracking. Four domain-specific specialist agents evaluate the case against established diagnostic criteria in parallel, and an arbitration agent synthesizes their opinions into a final classification. \textbf{(3)}, Report generation: AD-CARE returns a staging rationale, four-class differential probabilities (NC/MCI/AD/non-AD), specialist opinion summaries, supporting evidence, MRI atrophy visualization, diagnostic skill outputs, and downloadable structured reports.}\label{web_page}
\end{figure}

\subsection{A web-based interface enables clinical deployment}\label{web_app}
To support clinical deployment, we implemented a web interface that mirrors the practical workflow of case entry, automated analysis, and structured report export (Fig.~\ref{web_page}). Users can enter demographic information, cognitive scores, behavioral and neuropsychiatric assessments, medical history, APOE status, and functional assessments in text form, and can optionally upload structural MRI when available.

The interface exposes the underlying plan--execute--aggregate workflow and returns a structured report containing the final four-class differential diagnosis, confidence, supporting and contradicting evidence, criteria consistency notes, and recommended next steps. Reports can be exported in PDF or Word format for documentation and downstream integration. This implementation demonstrates that the AD-CARE framework can be operationalized in a clinician-facing environment rather than remaining only a retrospective benchmark model.

\section{Discussion}\label{sec3}
The central motivation of this work is to test whether a criteria-guided LLM agent framework for Alzheimer's disease dementia assessment can operate across clinical environments where patient populations, routine data availability, and computational resources differ substantially. Our findings indicate that AD-CARE meets several practical requirements for translation, including robustness across cohorts, more equitable subgroup behavior, measurable reader-study utility, and portability across backbone models.

\subsection{Robustness Across Diverse Clinical Settings}
A fundamental challenge for medical AI is achieving stable performance across diverse populations, imaging protocols, and healthcare environments. AD-CARE was intentionally designed around this requirement. By evaluating the agent across three public cohorts and two independent in-house clinical cohorts, we show that AD-CARE maintains robust diagnostic accuracy despite substantial heterogeneity in demographics, scanner vendors, acquisition parameters, and cognitive assessment protocols. Notably, the clearest separation did not arise on the easiest conventional AD-focused benchmarks, but on NACC and the independent in-house cohorts, where diagnostic heterogeneity is greater and the four-class differential task is more demanding. This pattern suggests that a criteria-guided differential reasoning scaffold becomes especially useful as cases more closely resemble the ambiguity of real-world memory-clinic assessment.

\subsection{Concentration of Gains in the Hardest Decisions}
The performance pattern across cohorts and classes suggests that AD-CARE does not provide a uniform uplift across all diagnostic decisions. Rather, its largest gains appear where clinical reasoning is hardest. AD-CARE preserved strong performance on conventional NC/MCI/AD cohorts, but its advantage became more pronounced in the more heterogeneous four-class setting and in independent in-house cohorts. The class-wise analyses sharpen this interpretation. When the four-class cohorts were evaluated under a three-class staging view (NC / MCI / Dementia), method-to-method differences were comparatively modest, indicating that the gain is not explained by superior syndromic staging. The mean gain over the baseline average then increased from NC (\(\Delta\)F1 = +0.029) to MCI (+0.063) and AD (+0.071), and was largest for non-AD (+0.133). In other words, the benefit was concentrated not in the easiest distinctions, but in dementia differentiation itself, especially when alternatives to AD had to be recognized explicitly under incomplete routine evidence.

A further concern is that stronger non-AD detection might simply reflect over-calling of alternative dementias. Our specificity analysis argues against this interpretation. On ADNI and AIBL, where the label space contains no true non-AD cases, AD-CARE's non-AD false-positive rates remained below 1\%. Together, these findings suggest that the framework improves sensitivity to clinically important alternatives without paying for that gain through indiscriminate non-AD labelling.

\subsection{Adaptability to Modality Availability}
Real-world AD dementia assessment rarely follows the fully multimodal paradigm assumed by research cohorts. Instead, data availability is opportunistic and depends heavily on local infrastructure, economics, and clinical pathways. AD-CARE directly addresses this constraint by conditioning on whichever routine clinical data are present, rather than relying on imputation, exclusion, or rigid architectural requirements. Across strata defined by the number of missing feature groups, AD-CARE retained the highest macro-F1 at every level, and its advantage over the baseline average widened from +0.078 to +0.111 as the record became sparser. Even in high-missingness cases (\(\geq 3\) missing feature groups), it maintained a macro-F1 of 0.652, compared with 0.620 for GPT-5.4-mini and 0.588 for MDAgents. This resilience to incomplete data is especially relevant in first-line memory-clinic settings, where routine work-ups are heterogeneous and confirmatory biomarkers are often unavailable or selectively ordered.

\subsection{Mechanistic Interpretation of the Performance Gain}
The ablation analyses support a specific interpretation of why the framework helps. Removing the stage-aware syndromic staging module produced by far the largest collapse in overall performance (\(\Delta\)macro-F1 = $-$0.462), indicating that explicit stage organization is the dominant structural component rather than a superficial prompt wrapper. Removing the dementia-only specialist layer caused the next-largest framework-level decline and sharply reduced non-AD detection (\(\Delta\)F1 = $-$0.130), consistent with the idea that much of the gain emerges after dementia-stage cases are routed to etiologic arbitration. At the tool level, the comparatively larger contributions of behavioral and vascular evidence than MRI to overall performance are also consistent with the first-line memory-clinic setting, in which differential diagnosis often depends on distributed routine clinical signals rather than imaging alone. Taken together, these results argue that AD-CARE's benefit does not arise from generic LLM uplift alone, but from separating syndrome-level staging from dementia-specific differential diagnosis within a criteria-guided scaffold that more closely mirrors clinical reasoning.

\subsection{Fairness and Equity in Diagnostic AI}

Our fairness analyses further underscore the importance of jointly considering average performance and subgroup consistency when designing diagnostic AI for dementia. Across pooled age, sex, race, and education strata, AD-CARE achieved the highest accuracy in all 12 subgroups while showing tighter subgroup-to-overall sensitivity disparities than GPT-4o. Its worst-subgroup accuracy also remained at or near the strongest subgroup performance reached by the best baselines. This combination is clinically important: models that perform well on average but degrade in very old adults or underrepresented populations risk amplifying existing disparities in dementia diagnosis. Our findings suggest that coupling explicit clinical reasoning, structured evidence aggregation, and criteria-guided prompting can contribute to more equitable behavior, although dedicated fairness interventions will still be needed in future iterations.

\subsection{Clinical Utility and Workflow Integration}
Beyond algorithmic performance, adoption in healthcare requires demonstrable improvements in clinical workflow. In a paired reader study at two hospitals, AD-CARE assistance increased pooled four-class accuracy from 50.0\% to 69.3\% and pairwise Cohen's $\kappa$ from 0.44 to 0.67, while reducing median reading time 1.48-fold. An exact McNemar test confirmed that corrections outnumbered regressions, $P = 1.9 \times 10^{-9}$. Gains were largest for non-AD recall, and high-disagreement cases accelerated most, by 1.86-fold. The seven-reader modal diagnosis moved from 21 of 40 to 31 of 40 correct cases, with no majority reversal from correct to incorrect. This pattern is consistent with the algorithmic results above: the scaffold is most useful in decisions that demand explicit evidence organization and differential reasoning. These findings indicate that AD-CARE can improve both the accuracy and the efficiency of human diagnostic decision-making.

\subsection{Deployment and Regulatory Adaptability}
Global deployment further requires adaptability to heterogeneous computational and regulatory environments. Many health systems cannot host large proprietary LLMs due to constraints in hardware, data governance, or cost. Our backbone analysis shows that AD-CARE reduces dependence on any single frontier-scale model: on NACC, wrapping each of five backbones (GPT-4o, GPT-5.4-mini, DeepSeek-V3.2, DeepSeek-R1, and LingShu-32B) with the AD-CARE scaffold improved diagnostic performance relative to the corresponding standalone LLM. This backbone-agnostic gain is important for equitable deployment, because institutions may need to choose models based on governance, local hosting, or cost constraints rather than raw benchmark leadership alone.

\subsection{Limitations}
Our study has several limitations. 
First, although we leveraged three public cohorts and two in-house cohorts, these primarily represent patients evaluated in research-oriented or tertiary-care environments. While our evaluation included direct four-class testing in cohorts with non-AD labels, the non-AD category remains heterogeneous and relatively sparse in the in-house datasets. Broader validation across richer and more evenly sampled non-AD dementia spectra will therefore be important.

Second, the framework was evaluated primarily on heterogeneous routine clinical data and did not rely on amyloid or tau PET, fluid biomarkers, or other confirmatory assays as core inputs. This is consistent with the first-line memory-clinic setting that motivates the study, but it means that we do not yet address biomarker-confirmatory diagnosis or mixed-pathology adjudication.

Third, the two in-house cohorts were modest in size and drawn from a single racial group, which may limit the representativeness and external validity of our findings in more diverse populations.

Fourth, while we conducted prespecified subgroup analyses by age, sex, race, and education, fairness in clinical AI is inherently multidimensional. We did not explicitly account for other important factors such as socioeconomic status, language, comorbidities, or health system characteristics, all of which can influence cognitive performance, imaging patterns, access to diagnostic investigations, and labelling practices.

Fifth, our reader study was a retrospective, controlled evaluation based on previously collected NACC cases and involved seven clinicians at two hospitals, with a fixed unaided-then-assisted order. This design may not fully capture the complexity of real-world clinical workflows, counterbalanced condition order, or long-term behavioural adaptation among clinicians.

Finally, although the NACC backbone analysis now includes an open-weight medical model (LingShu-32B), most paired comparisons still rely on API-hosted models. This remaining dependence may pose challenges for deployment in privacy-sensitive or highly regulated healthcare environments, and future work should further test locally hosted open-source backbones.

\subsection{Future Directions}
Future work should therefore extend AD-CARE along several axes. Methodologically, incorporating additional modalities (e.g., tau PET, fluid biomarkers, wearable or digital biomarkers), longitudinal trajectories, and survival or progression endpoints could enable unified diagnostic--prognostic agents that assist across the entire disease course. Clinically, expanding the label space to include more finely resolved non-AD dementias and ambiguous or mixed-pathology presentations would better reflect real diagnostic dilemmas. From a systems perspective, integrating AD-CARE with electronic health records, testing multilingual deployments, and conducting prospective, multi-center implementation studies will be essential to assess usability, trust, and impact on care delivery. More broadly, our findings suggest that clinically aligned, criteria-guided LLM agents can provide a scalable template for decision support in other neurodegenerative and cognitive disorders, provided that similar attention is paid to real-world heterogeneity, fairness, and human--AI collaboration.

In summary, we show that a criteria-guided LLM agent framework can meet several key requirements for real-world Alzheimer's disease dementia assessment. AD-CARE achieves robust performance across heterogeneous international cohorts, remains stable under substantial missing-modality patterns, and demonstrates comparatively equitable behavior across demographic subgroups. Its gains are concentrated in the most clinically difficult parts of the task, especially dementia differentiation under incomplete routine evidence, and the ablation results indicate that these gains are driven by stage-aware routing and dementia-specific arbitration rather than generic LLM scaling alone. In a paired reader study, the agent improved clinicians' diagnostic accuracy and efficiency, including on the cases that readers otherwise disagreed about. Importantly, AD-CARE maintains strong performance across different backbone models, supporting its feasibility for deployment in diverse healthcare environments. These findings position agentic systems as a promising path toward scalable, equitable, and deployable decision-support tools in dementia care.

\section{Methods}\label{method_part}
\subsection{Inclusion and Exclusion Criteria}
The study qualified for exemption from review by the local institutional review board because all neuroimaging and clinical data were received in de-identified form from external study centers, which confirmed adherence to ethical standards and documented informed consent from all participants ([2025]286-002 for XWH, II2026-115-01 for SYSUH). No compensation was provided to participants. Case labels were taken directly from the clinical diagnoses assigned by each source cohort. We did not retrospectively reclassify MCI by presumed etiology, because the goal was to preserve the diagnostic granularity and uncertainty of routine memory-clinic assessment. For the four-class setting, dementia attributed primarily to Alzheimer's disease was labeled AD, whereas available alternative dementia diagnoses were consolidated into a non-AD category.

Because AD-CARE was evaluated without model training and all data were used exclusively for testing, repeated visits from the same participant were treated as independent cases and were not considered information leakage. Across cohorts, eligible cases required (i) a documented clinical diagnosis, (ii) core demographic information (age, sex, and years of education), and (iii) a volumetric 3D T1-weighted MRI acquired within three months of the recorded diagnosis. When multiple MRIs and diagnostic entries were available within this interval, the MRI closest in time to the diagnosis was selected. Additional routine clinical variables, including neuropsychological testing, functional assessments, medical and psychiatric history, behavioral symptoms, biospecimen results, and genetic information, were incorporated whenever available; their absence did not trigger case exclusion. Supplementary Fig.~1 summarizes the cohort-specific inclusion workflow before and after application of these criteria.


\subsection{Clinical Data Representation}
AD-CARE was designed for first-line dementia assessment from \emph{heterogeneous routine clinical data}, rather than from biomarker-complete research panels. We formalized each case \(i\) as a partially observed clinical evidence set
\begin{equation}
    \mathbf{x}_i = \left\{x_i^{(m)}\right\}_{m=1}^{M}, \qquad
    \Omega_i = \left\{m \; | \; x_i^{(m)} \text{ is observed}\right\},
\end{equation}
where \(x_i^{(m)}\) denotes the \(m\)-th available clinical variable or modality and \(\Omega_i\) is the case-specific observation set. AD-CARE operated directly on \(\mathbf{x}_i^{\Omega_i}\), that is, only on the subset of evidence actually present for that case.

The observed data were then transformed by a deterministic evidence organizer
\begin{equation}
    \mathbf{s}_i = F_{\mathrm{org}}\!\left(\mathbf{x}_i^{\Omega_i}\right),
\end{equation}
where \(\mathbf{s}_i\) is a structured patient summary with predefined sections covering demographics, cognitive assessment, functional status, behavioral and neuropsychiatric symptoms, mood, motor and neurological findings, medical history, psychiatric history, physical examination, family history, genetic information, and other clinical data. This step was rule-based and did not rely on an LLM, thereby reducing variability in evidence presentation across cases.

Importantly, AD-CARE did not impute missing modalities or synthesize unobserved findings. Variables absent from \(\Omega_i\) simply remained absent from \(\mathbf{s}_i\), and downstream diagnostic skills returned partial or unavailable outputs when their required inputs were missing. This design preserved the clinical realism of routine work-ups while providing a stable and reproducible evidence bundle for subsequent reasoning.

\subsection{Criteria-guided Differential Reasoning Workflow}
As noted in Section~\ref{system_overview}, AD-CARE is a criteria-guided LLM agent framework for Alzheimer's disease dementia assessment from heterogeneous routine clinical data. Rather than mapping all inputs directly to a flat end-point classifier, the framework is organized around a \emph{criteria-guided differential reasoning scaffold} (Fig.~\ref{overview}b) with four clinically ordered stages: evidence organization and skill extraction, stage-aware syndromic staging, dementia-only specialist differentiation, and final etiologic arbitration. This architecture was designed to mirror memory-clinic reasoning, in which clinicians first organize the available evidence, then establish syndrome stage, and only thereafter adjudicate dementia etiology.

\noindent \textbf{Stage 1: Evidence organization and skill extraction.} In parallel with summary construction, AD-CARE invoked six tool-driven diagnostic skills over their corresponding observable subsets:
\begin{equation}
    \mathbf{z}_i^{(k)} = T_k\!\left(\mathbf{x}_i^{\Omega_i \cap \mathcal{M}_k}\right),
    \qquad k = 1,\dots,6,
\end{equation}
where \(T_k\) is the \(k\)-th diagnostic skill and \(\mathcal{M}_k\) denotes the clinical variable set required by that skill. The overall structured evidence passed downstream was therefore
\begin{equation}
    \mathcal{E}_i = \left(\mathbf{s}_i, \mathbf{Z}_i\right), \qquad
    \mathbf{Z}_i = \left\{\mathbf{z}_i^{(1)}, \dots, \mathbf{z}_i^{(6)}\right\}.
\end{equation}
Unlike generic agent pipelines that rely on free-form planning before domain grounding, AD-CARE introduced the LLM only after the case had been transformed into a clinically organized evidence bundle.

\noindent \textbf{Stage 2: Stage-aware syndromic staging.} The organized case summary and structured skill outputs were passed to an LLM-based staging module
\begin{equation}
    \left(\hat{y}_i^{\mathrm{stage}}, c_i^{\mathrm{stage}}, \mathbf{e}_i^{\mathrm{stage}}\right)
    = G_{\mathrm{stage}}\!\left(\mathcal{E}_i\right),
    \qquad
    \hat{y}_i^{\mathrm{stage}} \in \{\mathrm{NC}, \mathrm{MCI}, \mathrm{Dementia}\}.
\end{equation}
This stage-separated design prevented the model from conflating early cognitive impairment with dementia-subtype attribution. Because the boundary between late MCI and early dementia can be unstable in routine data, we added a deterministic safeguard,
\begin{equation}
    \tilde{y}_i^{\mathrm{stage}} =
    \begin{cases}
        \mathrm{Dementia}, & \hat{y}_i^{\mathrm{stage}}=\mathrm{MCI} \;\land\; \delta_i^{\mathrm{esc}} = 1, \\
        \hat{y}_i^{\mathrm{stage}}, & \text{otherwise},
    \end{cases}
\end{equation}
where \(\delta_i^{\mathrm{esc}}\) is a rule-based escalation flag triggered when structured functional evidence and global cognitive screening jointly indicate dementia-level impairment. This safeguard reduced the chance that borderline cases with substantial objective impairment would be prematurely terminated as MCI.

\noindent \textbf{Stage 3: Dementia-only specialist differentiation.} Only cases with \(\tilde{y}_i^{\mathrm{stage}}=\mathrm{Dementia}\) proceeded to etiologic assessment. For these cases, AD-CARE instantiated four explicit disease-oriented specialist modules
\begin{equation}
    o_i^{(d)} = H_d\!\left(\mathcal{E}_i; \mathcal{C}_d\right),
    \qquad
    d \in \mathcal{D} = \{\mathrm{AD}, \mathrm{LBD}, \mathrm{FTD}, \mathrm{VaD}\},
\end{equation}
where \(\mathcal{C}_d\) denotes the disease-specific criteria and prompt constraints associated with each diagnostic hypothesis~\cite{mckeith2017diagnosis,mckhann2011diagnosis,sachdev2014diagnostic,rascovsky2011sensitivity}. Less prevalent non-AD etiologies, such as NPH or PRD, were not instantiated as standalone specialist agents in the current implementation, but could still contribute to the final non-AD decision through the shared evidence bundle and arbitration stage. Each specialist returned a structured opinion about the degree of support for that syndrome.

\noindent \textbf{Stage 4: Final arbitration and structured output.} A final arbitration module reconciled the specialist opinions and generated the etiologic decision
\begin{equation}
    \left(\hat{y}_i^{\mathrm{eti}}, c_i^{\mathrm{eti}}, \mathbf{e}_i^{\mathrm{eti}}, u_i\right)
    = A\!\left(\mathcal{E}_i, \left\{o_i^{(d)}\right\}_{d \in \mathcal{D}}\right),
    \qquad
    \hat{y}_i^{\mathrm{eti}} \in \{\mathrm{AD}, \mathrm{non\mbox{-}AD}\},
\end{equation}
where \(c_i^{\mathrm{eti}}\) is the etiologic confidence estimate, \(\mathbf{e}_i^{\mathrm{eti}}\) is the corresponding evidence summary, and \(u_i\) denotes optional non-AD subtype suspicion. The final four-class prediction and returned explanation were then obtained by routing according to stage
\begin{equation}
    \left(\hat{y}_i, c_i, \mathbf{e}_i\right) =
    \begin{cases}
        \left(\tilde{y}_i^{\mathrm{stage}}, c_i^{\mathrm{stage}}, \mathbf{e}_i^{\mathrm{stage}}\right), & \tilde{y}_i^{\mathrm{stage}} \in \{\mathrm{NC}, \mathrm{MCI}\}, \\
        \left(\hat{y}_i^{\mathrm{eti}}, c_i^{\mathrm{eti}}, \mathbf{e}_i^{\mathrm{eti}}\right), & \tilde{y}_i^{\mathrm{stage}} = \mathrm{Dementia}.
    \end{cases}
\end{equation}
Accordingly, the structured report returned by AD-CARE can be written as
\begin{equation}
    \mathcal{R}_i = \left(\hat{y}_i, c_i, \mathbf{e}_i, \mathbf{Z}_i, u_i\right),
\end{equation}
with \(u_i = \varnothing\) for routed NC/MCI cases and for AD outputs, and \(u_i \in \{\mathrm{LBD}, \mathrm{FTD}, \mathrm{VaD}\}\) only when the final label is non-AD. 
All LLM interactions were constrained to structured JSON outputs with automatic retry when malformed responses occurred, so that downstream processing remained deterministic and auditable.

\subsection{Tool-driven Diagnostic Skills}\label{tool_description}
AD-CARE used six tool-driven diagnostic skills to convert heterogeneous routine inputs into clinically interpretable intermediate evidence. These skills were not independent predictors; rather, they functioned as modular evidence operators that transformed domain-specific subsets of \(\mathbf{x}_i\) into reasoning-ready outputs for the staging and arbitration modules. In implementation, these corresponded to Tool 1--6 for cognitive-functional, behavioral-neuropsychiatric, motor-neurological, vascular, MRI-atrophy, and genetic risk analysis, respectively. In general, each skill produced
\begin{equation}
    \mathbf{z}_i^{(k)} =
    \left(\mathbf{f}_i^{(k)}, \ell_i^{(k)}, \rho_i^{(k)}\right),
\end{equation}
where \(\mathbf{f}_i^{(k)}\) denotes structured findings, \(\ell_i^{(k)}\) denotes syndrome-oriented support or flag levels, and \(\rho_i^{(k)}\) denotes data completeness.

\noindent \textbf{Cognitive-functional skill \(T_{\mathrm{cog}}\):} A deterministic scoring engine over MMSE, MoCA, functional status, and domain-level neuropsychological tests. This module harmonizes equivalent measurements across UDS2 and UDS3-style batteries when only one version is available, estimates cognitive severity, infers impaired domains such as memory, executive function, language, and attention, and quantifies the functional consequence of cognitive decline. Its outputs primarily support the syndromic separation of NC, MCI, and Dementia.

\noindent \textbf{Behavioral-neuropsychiatric skill \(T_{\mathrm{beh}}\):} A rule-based symptom extraction module over neuropsychiatric inventory items, severity scores, REM-sleep-related history, and depression-related fields. It produces disease-oriented behavioral flags, including DLB-supporting features such as hallucinations and REM-sleep-behavior-disorder-related findings, as well as FTD-supporting features such as disinhibition, apathy, aberrant motor behavior, and appetite change.

\noindent \textbf{Motor-neurological skill \(T_{\mathrm{mot}}\):} A neurological feature parser over parkinsonian signs, gait disturbance, focal neurological deficits, cortical signs, and related movement-disorder history. It returns structured evidence on parkinsonism, DLB-oriented motor support, VaD-related neurological signs, and other non-AD clues relevant to etiologic arbitration.

\noindent \textbf{Vascular skill \(T_{\mathrm{vas}}\):} A vascular burden summarizer over stroke or transient ischemic attack history, multiplicity of cerebrovascular events, vascular risk factors, blood-pressure readings, and small-vessel-disease indicators. It generates graded vascular support signals for assessing whether cerebrovascular burden plausibly contributes to cognitive impairment.

\noindent \textbf{MRI-atrophy skill \(T_{\mathrm{mri}}\):} A tool-driven imaging subpipeline built on FastSurfer regional volumetry~\cite{henschel2020fastsurfer}. For each region \(r\), intracranial-volume-normalized measurements were converted into age- and sex-referenced normative deviations
\begin{equation}
    z_{i,r} = \frac{v_{i,r} - \mu_{b(i),r}}{\sigma_{b(i),r}},
\end{equation}
where \(v_{i,r}\) is the normalized regional volume for case \(i\), and \(\mu_{b(i),r}\) and \(\sigma_{b(i),r}\) are the mean and standard deviation of the corresponding age-sex reference bin \(b(i)\). Hemispheric asymmetry for lobe \(\ell\) was quantified as
\begin{equation}
    a_{i,\ell} = \frac{\left|v_{i,\ell}^{L} - v_{i,\ell}^{R}\right|}{\left(v_{i,\ell}^{L} + v_{i,\ell}^{R}\right)/2}.
\end{equation}
These regional deviations and asymmetry indices were then mapped to clinically recognizable atrophy patterns such as medial-temporal-predominant, frontal-predominant, posterior cortical, diffuse, asymmetric, or minimal atrophy.

\noindent \textbf{Genetic risk skill \(T_{\mathrm{gen}}\):} A genotype interpretation tool that maps APOE status to ordinal AD-support levels. Genetic information is treated as probabilistic support rather than a deterministic label source, so APOE4 carriage increases support for AD but does not exclude non-AD dementia.
\subsection{Baseline Methods}
We benchmarked AD-CARE against medical LLMs, general-purpose LLMs, reasoning-oriented LLMs, and a general medical agent framework.

\noindent \textbf{MedGemma-4B}~\cite{sellergren2025medgemma}: A medical-domain language model used to benchmark compact, domain-specialized LLM performance on the same diagnostic prompts as AD-CARE.

\noindent \textbf{HuluMed-7B}~\cite{jiang2025hulu}: A lightweight medical LLM included to represent smaller clinically oriented open models.

\noindent \textbf{LingShu-32B}~\cite{xu2025lingshu}: A larger medical LLM used to assess whether domain pretraining alone can match the proposed agentic scaffold.

\noindent \textbf{DeepSeek-V3.2}~\cite{deepseekv3technicalreport}: A general-purpose LLM used as a strong open general baseline.

\noindent \textbf{GPT-4o}~\cite{gpt-4o}: A general-purpose multimodal LLM used both as a standalone baseline and as the default backbone for AD-CARE in the primary experiments.

\noindent \textbf{DeepSeek-R1}~\cite{guo2025deepseek_r1}: A reasoning-oriented LLM included to test whether generic chain-of-thought capability alone can solve the differential diagnosis task.

\noindent \textbf{GPT-5.4-mini}~\cite{openai2026gpt54mini}: A compact reasoning-oriented model used to benchmark strong general reasoning performance under the same prompting setting.

\noindent \textbf{MDAgents}~\cite{kim2024mdagents}: A general-purpose medical agent framework instantiated with a GPT-4o backbone, included as the primary agentic baseline. In the main-text analyses, we focus on the strongest five representative baselines (DeepSeek-V3.2, GPT-4o, DeepSeek-R1, GPT-5.4-mini, and MDAgents), while full-panel results for all methods are reported in Supplementary Table~4.

\subsection{Reader Study Validation}\label{expert-level validation}
We evaluated the clinical utility of AD-CARE in a paired reader study with seven clinicians from Xuanwu Hospital, Capital Medical University (XWH; $n=3$) and the Third Affiliated Hospital of Sun Yat-sen University (SYSUH; $n=4$). Readers independently reviewed 40 cases drawn from NACC, balanced across the four diagnostic classes used throughout this study (NC, MCI, AD, and non-AD dementia; 10 each) and across MRI availability (20 with T1-weighted MRI and 20 without). Each case included the same class of routine clinical variables available to AD-CARE (demographics, cognitive and functional assessments, behavioural and neuropsychiatric measures, medical history, and APOE status when recorded). Readers assigned one of the four labels NC, MCI, AD, or non-AD.

Cases were stratified rather than drawn as a prevalence sample. The 40-case set comprised 24 advantage cases (AD-CARE correct, GPT-4o backbone incorrect), 8 control cases (both correct), and 8 AD-CARE-wrong cases (AD-CARE incorrect), equally allocated across classes and MRI availability (seed 20260710). Standalone AD-CARE accuracy on this set was 80\%. This enrichment tests whether the algorithmic advantage can be taken up in human decisions; it does not reproduce outpatient prevalence, and the 80\% figure should not be read as clinic-wide agent accuracy.

The study used a two-session within-reader design with a fixed condition order. In session 1 (unaided), readers reviewed the raw clinical record without agent output. In session 2 (assisted), they reviewed the same cases together with AD-CARE's structured report, including the diagnostic impression, supporting evidence, and specialist summaries. A washout of 7--20 days separated the two sessions (median 10 days). Case order was randomized independently within each session. During the assisted session, readers were instructed to treat the agent output as decision support only and to retain full responsibility for the final diagnosis; they could accept, modify, or reject the suggestion. All readings were performed individually without discussion among readers. Per-case reading time was defined as the interval from starting a case to submitting a diagnosis; times exceeding 1,000~s (abandoned page-open events) were retained for medians, which are insensitive to these events, and were excluded from means reported in the Supplementary Information. Further protocol notes are provided in Supplementary Note~S4.

\subsection{Statistical Analysis}
\subsection{Statistical Analysis}
We used one-way analysis of variance (ANOVA) for continuous variables and the $\chi^{2}$ test for categorical variables to assess differences in population characteristics across the diagnostic groups available within each cohort (NC, MCI, AD, and non-AD when present). All hypothesis tests were two-sided with a significance threshold of 0.05. Algorithmic diagnostic performance was summarized using overall accuracy, mean class accuracy (mACC; the unweighted mean of per-class recall), and macro-averaged F1. For each metric we report the point estimate together with a 95\% confidence interval from a non-parametric bootstrap with 2,000 case-level resamples~\cite{CI}. Subgroup analyses used the same metrics within each demographic stratum.

In the reader study, the primary comparison was paired diagnostic correctness across 280 reader--case observations (seven readers $\times$ 40 cases), tested with an exact McNemar test. Overall accuracy, macro-F1, and class-wise recall were computed on the pooled reads. Reading times are reported as medians. Inter-reader agreement was summarized as pairwise percent agreement and as Cohen's $\kappa$ averaged across reader pairs. The panel diagnosis was the plurality label among the seven readers; a case was majority-correct when this label matched ground truth. Case difficulty was defined from unaided pairwise agreement, ranked into tertiles of 13, 14, and 13 cases. The association between unaided agreement and unaided case-level accuracy was summarized with Spearman's rank correlation. Follow-through was defined as identity between the assisted diagnosis and the AD-CARE prediction for that case. 



\bibliography{sn-bibliography}
\section{Data Availability}
Data from ADNI, AIBL can be downloaded from the LONI website at https://ida.loni.usc.edu. NACC data can be downloaded at https://naccdata.org. Restrictions apply to the availability of the raw in-house data, which was used with institutional permission through ethics approval for the current study, and are thus not publicly available. Please email all requests for academic use of raw and processed data to the corresponding author. All requests will be evaluated based on institutional and departmental policies to determine whether the data requested is subject to intellectual property or patient privacy obligations. Data can only be shared for non-commercial academic purposes and will require a formal material transfer agreement.
\section{Code Availability}
Python scripts as well as help files along with information on the study population are made available on GitHub (https://github.com/SDH-Lab/AD-CARE). All prompts used in this work are in the supplementary file.

\section{Acknowledgments}
The authors also acknowledge the use of AI tools for assistance in grammar enhancement and spelling checks during the preparation of this manuscript.

\section{Author Contributions}
Y.C., S.Y., X.M., and S.W. are the corresponding authors. W.H designed, developed, and validated the agent framework. G.Y., S.B., S.Y., H.X., Y.X., X.M., Y.C., J.L. and L.Long performed data collection. N.Y. and W.H performed statistical analysis. W.H., J.W. and Y.F. generated the figures and tables. S.Y., X.M. and S.B. completed the reader study. Y.D., L.Liu, S.W. provided guidance on the modeling framework. W.H., L.Liu, S.W. wrote the manuscript. L.Liu, S.W., N.M., M.Y., S.Y., Q.N., S.L., Z. K., and K.T. revised the manuscript. All authors approved the final version for publication and agreed to be accountable for all aspects of the work, ensuring that questions related to the accuracy or integrity of any part of the work are appropriately investigated and resolved.
\section{Competing Interests}
We declare that the authors have no competing interests as defined by Nature Portfolio, or other interests that might be perceived to influence the results and/or discussion reported in this paper.

\end{document}